\documentclass[12pt]{article}
\usepackage[T1]{fontenc}
\DeclareUnicodeCharacter{0229}{\c{e}}

\usepackage[font=scriptsize,labelfont=bf]{caption}
\usepackage{setspace}

\usepackage[compact]{titlesec}
\titlespacing{\section}{%
  0pt}{%              left margin
  0em}{% space before (vertical)
  0pt}% 
\titlespacing{\subsection}{%
  0pt}{%              left margin
  0em}{% space before (vertical)
  0pt}% 
\titlespacing{\subsubsection}{%
  0pt}{%              left margin
  0em}{% space before (vertical)
  0pt}% 
\titlespacing{\paragraph}{%
  0pt}{%              left margin
  0em}{% space before (vertical)
  0.5em}% 

\usepackage{amsmath,amssymb}
\usepackage{graphicx,psfrag,epsf}
\usepackage{subfig}

\usepackage{enumerate}

\usepackage{enumitem}

\usepackage{natbib}
\usepackage{soul}
\usepackage[normalem]{ulem}
\usepackage{lscape}

\usepackage{float,multirow,booktabs}
\usepackage[table,xcdraw]{xcolor}
\usepackage{url}

\usepackage{multicol}
\usepackage[section]{placeins}
\usepackage{rotating,titlesec}
\usepackage[english]{babel}

\usepackage[utf8]{inputenc}

\usepackage{algorithm,amsmath,tabularx}
\usepackage[noend]{algpseudocode}
\usepackage{hyperref}

\usepackage{bm}
\usepackage{nccmath}
\usepackage{multirow}

\hypersetup{
  colorlinks=true,
  allcolors=blue
}

\usepackage{tcolorbox}
\usepackage{cleveref}

\newcounter{algsubstate}

\makeatletter
\newcommand{\multiline}[1]{%
	\begin{tabularx}{\dimexpr\linewidth-\ALG@thistlm}[t]{@{}X@{}}
		#1
	\end{tabularx}
}
\makeatother

\begin{document}
\setstcolor{red}

\def\spacingset#1{\renewcommand{\baselinestretch}%
    {#1}\small\normalsize} \spacingset{1}

%%%%%%%%%%%%%%%%%%%%%%%%%%%%%%%%%%%%%%%%%%%%%%%%%%%%%%%%%%%%%%%%%%%%%%%%%%
% TO DELETE
\newcommand{\tempsupref}[1]{\textcolor{red}{\textsf{[#1]}}}
\newcommand{\vb}[1]{\textcolor{magenta}{\textsf{[#1]}}}
\newcommand{\redacted}[1]{\textcolor{red}{\textsf{[#1]}}}
\newcommand{\jk}[1]{\textcolor{violet}{\textsf{[JK: #1]}}}
\newcommand{\jkedit}[1]{
\textcolor{violet}{#1}
}
%%%%%%%%%%%%%%%%%%%%%%%%%%%%%%%%%%%%%%%%%%%%%%%%%%%%%%%%%%%%%%%%%%%%%%%%%%
	
%%%%%%%%%%%%%%%%%%%%%%%%%%%%%%%%%%%%%%%%%%%%%%%%%%%%%%%%%%%%%%%%%%%%%%%%%%%%%%
	
% 	\spacingset{1.32} % DON'T change the spacing!
	
% 	\title{\bf Functional Integrative Bayesian Analysis of High-dimensional Multiplatform Genomic Data}
	
\title{\Large \bf Spatially Structured Regression for Non-conformable Spaces: Integrating Pathology Imaging and Genomics Data in Cancer
}
%Dual Random Effect and Main Effect Spatial Selection for Pathology Imaging and Genomics Data}

\author{Nathaniel Osher$^{a, *}$, Jian Kang$^a$, Arvind Rao$^{b,c,a}$, Veerabhadran Baladandayuthapani$^a$ \\~\\
$^{a}${\small Department of Biostatistics, University of Michigan, Ann Arbor, MI 48105}\\
$^{b}${\small Department of Computational Medicine \& Bioinformatics, University of Michigan, Ann Arbor, MI}\\
$^{c}${\small Department of Radiation Oncology, University of Michigan, Ann Arbor, MI}\\
$^{*}${\small Corresponding author. E-mail: \href{mailto:oshern@umich.edu}{oshern@umich.edu}}}

\date{\today}

\maketitle
\setcounter{page}{0}
\thispagestyle{empty}

\begin{abstract}

The spatial composition and cellular heterogeneity of the tumor microenvironment plays a critical role in cancer development and progression. High-definition pathology imaging of tumor biopsies provide a high-resolution view of the spatial organization of different types of cells. This allows for systematic assessment of intra- and inter-patient spatial cellular interactions and heterogeneity by integrating accompanying patient-level genomics data. However, joint modeling across tumor biopsies presents unique challenges due to non-conformability (lack of a common spatial domain across biopsies) as well as high-dimensionality. To address this problem, we propose the \underline{D}ual \underline{r}andom \underline{e}ffect and \underline{m}ain \underline{e}ffect selection model for \underline{Spa}tially \underline{s}tructured r\underline{e}gression model (\texttt{DreameSpase}). \texttt{DreameSpase} employs a  Bayesian variable selection framework that facilitates the assessment of spatial heterogeneity with respect to covariates both within (through fixed effects) and between spaces (through spatial random effects) for non-conformable spatial domains.  We demonstrate the efficacy of \texttt{DreameSpase} via simulations and integrative analyses of pathology imaging and gene expression data obtained from $335$ melanoma biopsies. Our findings confirm several existing relationships, e.g. neutrophil genes being associated with both inter- and intra-patient spatial heterogeneity, as well as discovering novel associations. We also provide freely available and computationally efficient software for implementing \texttt{DreameSpase}. 
 
\end{abstract}

\noindent%

\vfill

\newpage

\spacingset{1.8}

\section{Introduction}\label{sec:introduction}

\subsection{Scientific motivation and background}

\paragraph{The tumor microenvironment} Tumors consist of many different components with complex interrelationships, including the vascular network within the tumor, the extracellular matrix of the tumor cells, and the immune cell composition of the tumor (\citealt{Sun16}). These structures are collectively referred to as the \textit{Tumor Microenvironment} (TME). The TME plays a critical role in the cancer development, progression and the efficacy of certain treatments (\citealt{Sadeghi21}). Among its various components, the immune composition of the TME 
has garnered much interest among cancer researchers. This interest is due to increasing recognition of cancer's ability to avoid destruction by the immune system as one of the organizing principles (``hallmarks'') of cancer evolution  (\citealt{Hanahan22}). Briefly, the immune composition of the TME is a fractally complex ecosystem. The broad class of ``immune cells'' is divided into a multitude of immune cell subtypes (\citealt{zoghi23chap1}). Sometimes subtypes of the same immune cell can even have divergent associations with prognosis (\citealt{Li21}). %; \citealt{Tan22}). 
Adding to this complexity is the variability in the spatial composition of the TME. While the prevalence of certain cell types within the TME can be informative, more recent research has shown that the relative spatial locations of different immune cell types (i.e. the tumor geography) offer insights into tumor classification and prognostic assessment (\citealt{Krishnan22}). 

\paragraph{Digital pathology} The complexity of TME has spurred research in the development and application of systematic quantitative assessment and computational modeling, an area often collectively referred to as \textit{Digital Pathology}. Digital pathology encompasses a broad array of aims and methods in the context of cancer pathology imaging, including classification of cells in the TME, determination of tumor grade, and the evaluation of spatial and morphological patterns in the tumors (\citealt{Heindl15}). Given the time and labor intensive nature of assessing the spatial composition of the TME, the focus on evaluating spatial and morphological patterns has become a central aim in digital pathology research (\citealt{Baxi22}). While there are many spatial characteristics of tumors that are potentially of interest, such as the distributions of stromal cells and blood vessels within the tumor, in this paper we focus on modeling tumor-immune interaction. Our goal is to characterize the spatial configurations of tumor cells and immune cells where the two cell types tend to cluster as positively interacting, and configurations where they tend to avoid one another as negatively interacting. This spatial tendency reflects the infiltration of the tumor by immune cells.

\paragraph{Motivating dataset and key scientific questions} 

\begin{figure}
    \centering
    \includegraphics[width=10cm]{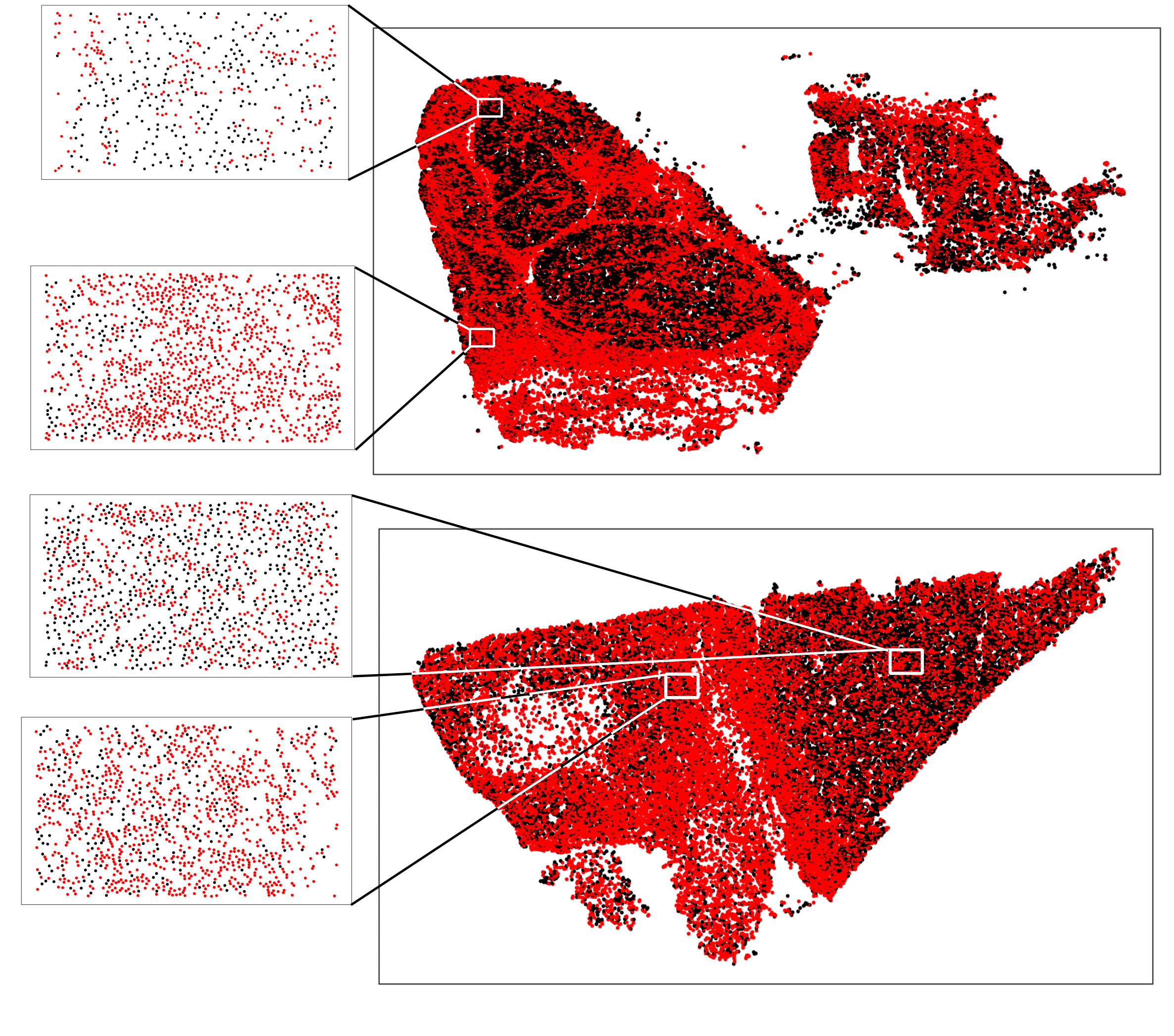}
    \caption{\textbf{Plot of two biopsies from motivating Melanoma data set}. Cell types and locations for two biopsies in the data set. Black points represent tumor cells, while red points represent immune cells. Enlarged portions are sub-regions with varying degrees of tumor-immune interaction. }
    \label{fig:heterogeneity}
\end{figure}

The motivating imaging data for this project comes from Melanoma biopsies collected by The Cancer Imaging Archive (\citealt{tcia13}). After some image preprocessing, the cell-type (i.e. phenotypes) and locations of individual cells in the biopsies are determined - see Figure \ref{fig:heterogeneity} for examples of two such biopsies (\citealt{Osher23}); additional details provided in Section \ref{sec:application}. In addition to the high definition pathology images, bulk RNA-seq data were collected for each patient at the biopsy level. This data naturally raises three fundamental scientific questions. First, what is the appropriate method of quantification of spatial tumor-immune interactions  within and between different biopsies (in other words, the extent to which immune cells interact with tumor cells)? 
Second, how does the variability in these interactions between biopsies relate to molecular composition of the tumor? Investigating this may shed light on the biological mechanisms most responsible for tumor-immune interactions. Third and finally, how does the variability in interaction \textit{within} biopsies (i.e. intra-tumoral heterogeneity) relate to relate to molecular composition of the tumor? Fundamentally, answering these three questions requires a synthesis of information from both imaging and genomics data. 

\subsection{Statistical challenges} 

\paragraph{Data structure} 
%It is common in practice to encounter data that has a spatially correlated structure. 
As demonstrated by Figure \ref{fig:heterogeneity}, tumor-immune cell organization can vary considerably both within and between biopsies, which can have drastically different physical structures. Furthermore, it is of interest how these cellular interactions vary both within and between biopsies with respect to a set of covariates of interest (i.e. gene expression data).  The resulting data are complex and multifaceted, yet three key aspects are particularly influential in determining the appropriate analysis approach. We refer to these aspects as the \textit{resolution of the outcome}, the \textit{resolution of the covariates} and the \textit{conformability of the spaces}. The \textit{resolution of the outcome} refers to the ``level'' at which the outcome of interest is observed (i.e. at the biopsy level or spatially across the biopsy), and the \textit{resolution of the covariates} is defined analogously. Conformability refers to whether or not the spaces share a similar structure.

A great deal of work has been done in modeling of spatial data from conformable spaces that can be ``mapped'' to a common space, along with high-resolution outcomes and covariates (outcomes and covariates that are observed at many spatial locations- see \citealt{MacNab22} and references there-in). Additionally, a considerable amount of work has been done on data from conformable spaces with either high-resolution outcomes and low-resolution covariates or low-resolution outcomes and high-resolution covariates, especially in the context of brain imaging data (\citealt{Bowman14} and references there-in). However, comparatively little work has been done to model spatial data from non-conformable spaces with high resolution outcomes and low resolution covariates. 

The structure of our motivating data set is considerably different than those listed above. Each biopsy is partitioned into non-overlapping sub-regions, which can then be modeled using spatial point processes. Measures of interactions can then be summarized for each of the  sub-regions, and modeled with respect to biopsy level covariates such as  gene expression data and/or patient demographics and clinical outcomes. The result is data with high resolution outcomes, low-resolution covariates, and non-conformable spaces. While such structure poses no great obstacle to modeling heterogeneity with respect to covariates \textit{between} patients, modeling spatial heterogeneity \textit{within} patients with respect to covariates requires considerably more thought.

\paragraph{Existing methods} While there are existing methods that model interactions in the TME, they largely focus on within biopsy modeling rather than joint modeling across biopsies. Most notably in the statistical domain there has been work to develop novel spatial models to assess interaction within the TME (\citealt{Li19}; \citealt{LiPotts19}). There have also been successful applications of machine learning methods to this end, focusing on the application of deep learning models and clustering (\citealt{Abousamra22}; \citealt{Saltz18}). To our knowledge, a joint model to borrow strength across biopsies that would answer the aforementioned scientific questions of interest has not been developed. Another solution to this problem is to convert the non-conformable spatial data to functional representations, which can then be analyzed in the paradigm of functional data analysis (\citealt{Vu22}; \citealt{Chervoneva21}; \citealt{Yang17}). While this is a viable course of action for cross-biopsy modeling, it does not 
take into account the inherent spatial structure and correlation between adjacent sub-regions.

\begin{figure}
    \centering
    \includegraphics[width=15cm]{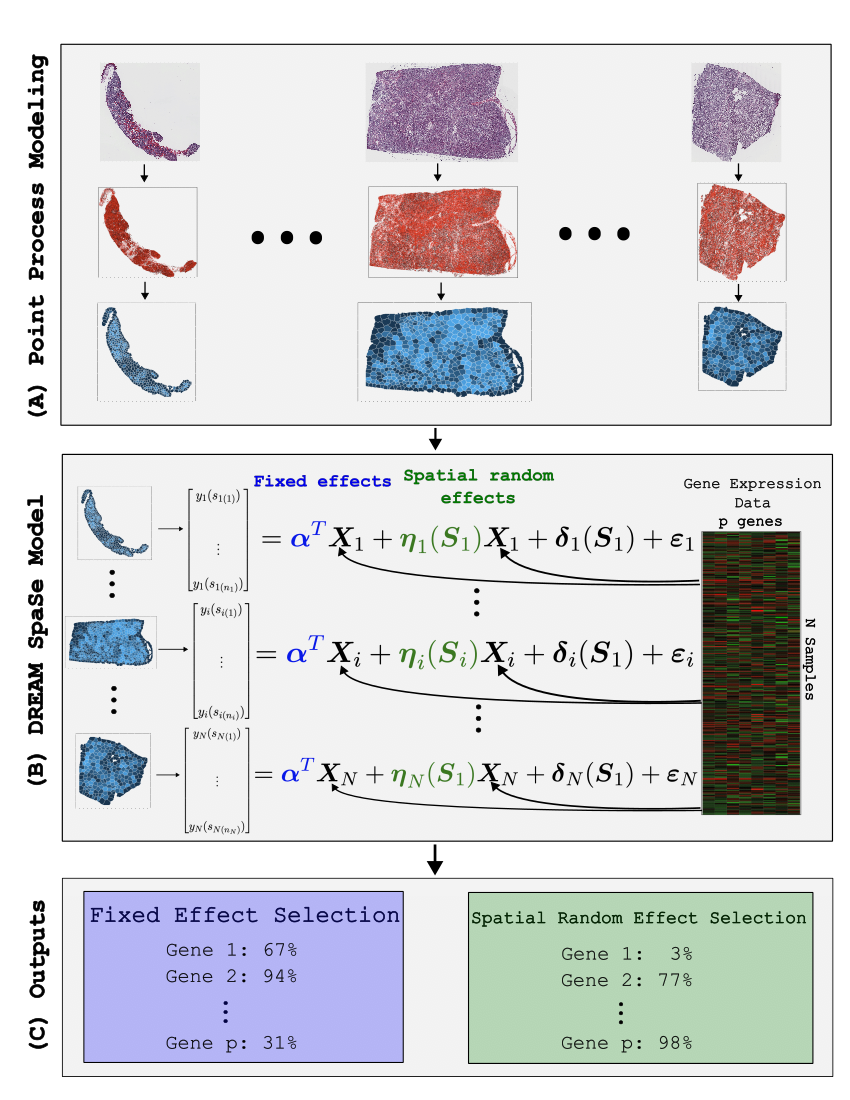}
    \caption{\textbf{Overview of the \texttt{DreameSpase} model.} \textbf{(A) Point process modeling}. Types and locations are determined for each cell in high definition biopsy images (top row to middle row). Biopsies are then divided into non-overlapping sub-regions, on which a measure of spatial interaction is computed (middle row to bottom row). \textbf{(B) \texttt{DreameSpase} model fitting.} Sub-regional outcomes are vectorized and the \texttt{DreameSpase} model is fit on the resulting vectors. Gene expression data, measured at the biopsy level, is used as the covariate set of interest. \textbf{(C) Fixed and Random Effect Selection.} Selection of both fixed and random effects are summarized for the model. fixed effects indicate association between covariates and average level of infiltration in a given biopsy, while random effects indicate association between covariates and heterogeneity of infiltration in a given biopsy.}
    \label{fig:modeloverview}
\end{figure}

\subsection{Model overview}\label{sec:modeloverview}

\paragraph{Model components} In this paper, we present the \underline{D}ual \underline{r}andom \underline{e}ffect and \underline{m}ain \underline{e}ffect selection model for \underline{Spa}tially \underline{s}tructured r\underline{e}gression (\texttt{DreameSpase}).  As conceptually outlined in Figure \ref{fig:modeloverview}, we first quantify the spatial interactions between tumor-immune cells through point process modeling  (Panel A). Subsequently, \texttt{DreameSpase} model extends the spike and slab prior traditionally used for fixed effects to the variance terms of spatial random effects. This allows for the joint modeling of inter-biopsy heterogeneity with respect to covariates via the fixed effects, and intra-biopsy heterogeneity with respect to covariates via the spatial random effects (Panel B). The novelty of the model lies in its ability to account for heterogeneity both within and between non-conformable spaces with respect to arbitrary covariates of interest. The model consists of two components of primary interest (Panel C): 

\begin{enumerate}
    \item \textit{Fixed effect selection component}. The model performs fixed effect selection via spike and slab priors. In the context of cancer genomics, this component accounts for the heterogeneity of the outcome of interest \textit{between} biopsies with respect to gene expression data. 

    \item \textit{Spatial random effect selection component}. In contrast to the fixed effect selection component, the random effect selection component accounts for the heterogeneity \textit{within} biopsies with respect to gene expression data. 
\end{enumerate}

As we demonstrate via simulation studies, gains in selection accuracy through joint modeling of the fixed and random effects are considerable as compared to non-spatial and independent (two-step) approaches. We apply \texttt{DreameSpase} to our motivating data set consisting of 335 images and associated gene expression data for various immune classes. We find that the neutrophil class was most consistently associated with within- and between-biopsy heterogeneity of interaction between immune cells and tumor cells. The association between several neutrophil class genes were consistent with the relationships previously established in the literature, while others were suggestive of potential novel relationships.

The remainder of the paper is organized as follows. Section \ref{sec:DreameSpase} details the formal model structure and formulation. Section \ref{sec:postinf} provides the details of the sampling algorithm and inferential summaries of interest. 
%offers a detailed derivation for the conditional distributions of the various components of the model. 
Section \ref{sec:simulationstudy} outlines the simulation studies and assessments with comparative methods. Section \ref{sec:application} presents the results of analyzing whole slide Melanoma biopsy pathology imaging and genomic data. Finally, we conclude with discussion, limitations, and possible future directions for research in Section \ref{sec:conclusionsandlimitations}.

\section{DreameSpase model}\label{sec:DreameSpase}

\subsection{Modeling tumor-immune cell interactions}\label{sec:datastructure}

\begin{figure}
    \centering
    \includegraphics[width=17cm]{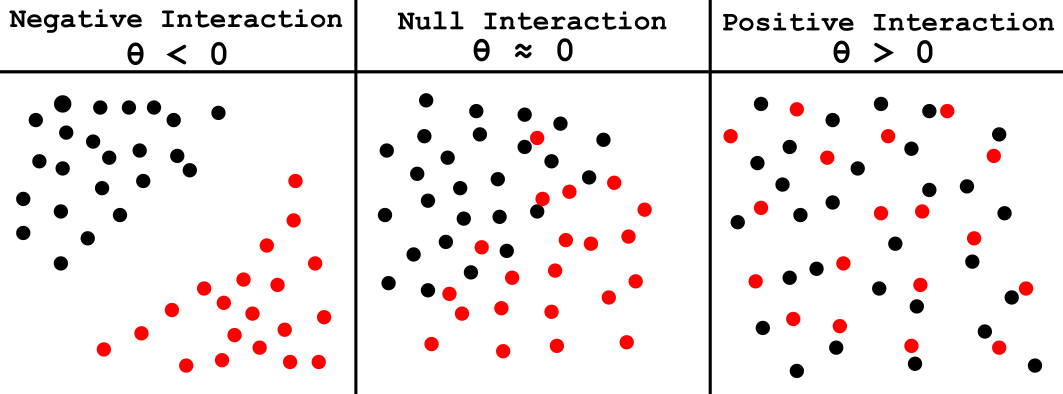}
    \caption{\textbf{Illustration of interaction/co-localization}. The leftmost panel illustrates low-interaction, i.e. the cells of different types do not tend to be near one another. The middle panel illustrates moderate interaction, i.e. some cells of both types tend to be near one another, while others do not. The third and final panel illustrates high interaction, i.e. the cells of both types exhibit a strong tendency to be near cells of the other type.}
    \label{fig:interactioncartoon} 
\end{figure}

\paragraph{The hierarchical Strauss model} In order to quantify the interaction between these two types of cells, we employ a Bayesian \textit{Hierarchical Strauss Model} (HSM) (\citealt{Hogmander99}). The HSM offers a parametric model that can capture both positive and negative interaction between points of different types, i.e. the tendency of the two types to either cluster towards one another or ``avoid'' one another. We chose the HSM for two primary reasons. First, it can capture both positive and negative interaction between different types of points, which the more standard Strauss model cannot accomplish. Second, the HSM captures the inherent biological nature of interaction between the tumor and the associated immune cells. The HSM is called ``hierarchical'' because the modeling of the second type of points (in our case, immune cells) is done conditionally on the locations of the first type of points (in our case, tumor cells). While the interaction between the tumor and immune cells is complex, as a first order approximation it is natural to assume as a first order approximation that the tumor precedes the immune response, and thus it makes intuitive sense to model the immune response conditional on the locations of the tumor cells. The density of the HSM model is as follows:
\vspace*{-10pt}
$$
f(\bm{x}_1, \bm{x}_2) \propto 
\exp\left\{n_1 \beta_1  + n_2 \beta_2 +
S_{R}(\bm{x}_1, \bm{x}_2) \theta\right\},
\vspace*{-10pt}
$$

where the parameters $\beta_1$ and $\beta_2$ capture the first order intensity of points for types 1 and 2, respectively (tumor and immune cells), with $n_1$ and $n_2$ representing the number of points for the corresponding type. Of key importance in this model is the interaction parameter, $\theta$, that captures the tendency of the different types of points to exhibit positive or negative interaction. To gain some intuition as to why this is the case, first note that $S_{R}(\bm{x}_1, \bm{x}_2)$ counts the number of pairs of points within radius $R$ of one another where one is of type 1 and the other is of type 2. Note that $R$ is not a parameter that is fit, but is instead chosen prior to the model fitting based on knowledge of the particular domain being modeled. Thus, when $\theta$ is negative it ``penalizing'' the density put on configurations where points of the two types tend to cluster. Conversely, such configurations being ``rewarded'' when $\theta$ is positive. 
%In fact, this intuition is more or less a satisfactory account of how to interpret the fitted value of an interaction parameter. 
In essence, $\theta$ offers a highly interpretable summary of the spatial interaction between the different types of points:

\begin{itemize}
    \item $\theta \in (-\infty, 0)$ indicates negative interaction (as illustrated by the leftmost panel of Figure \ref{fig:interactioncartoon}), and lower values indicate more negative interaction.
    \item $\theta \in (0, \infty)$ indicates positive interaction (as illustrated by the rightmost panel of Figure \ref{fig:interactioncartoon}), and larger values indicate more positive interaction.
    \item $\theta = 0$ indicates null interaction, i.e. the points are ``indifferent'' to points of the other type, as illustrated by the middle panel of Figure \ref{fig:interactioncartoon}.
\end{itemize}

\paragraph{Biopsy partitioning} Whole slide high-resolution pathology images of biopsies typically contain between $10^4 - 10^5$ cells. For example, in our motivating data set, the median number of cells is $\sim50,000$, with the maximum across all biopsies $\sim190,000$. This volume means that simply computing a spatial summary on the biopsy level is challenging. Due to the sheer size of the biopsy any given spatial feature will tend to vary across the biopsy in ways that may be important. We are primarily interested in modeling interaction (the tendency of different cells to be near each other), which is quantified on a fairly local scale ($\sim 30 \mu m$) which is dictated by radius of influence of a single cell. In order to capture the heterogeneity of this interaction, we partition the biopsy into sub-regions defined by the locations of the individual tumor cells. In addition, this partitioning has the added benefit of improving the computational requirements of the HSM.

More specifically, let $N$ be the number of biopsies, respectively denoted $B_1, \dots, B_N$, each $B_i \subset \mathbb{R}^2$ is further partitioned into $n_i$ disjoint sub-regions $b_{i(1)}, \dots, b_{i(n_i)}$ with $b_{i(j)} \subseteq B_i$, $b_{i(j)} \cap b_{i(k)} = \varnothing$ for $k \neq j$, and $B_i = \bigcup\limits_{j=1}^{n_i} b_{i(j)}$. On sub-region $b_{i(j)}$ we observe a marked point process model $\bm{P}_{i(j)}$ consisting of two types of points, tumor cell and immune cell. 

The HSM can be fit on each of the non-overlapping sub-regions to determine the local level of tumor-immune interaction, as shown in panel A of Figure \ref{fig:modeloverview}. For more details on the HSM and the $\theta$ parameter, see Sections S1.2 and S1.3 of the supplementary materials. The model can be fit using standard Bayesian posterior sampling methods on the so-called pseudolikelihood function used in the estimation of these models; see Section S1.2 of the supplementary materials for model fitting details. While there are many options in the spatial statistical literature for measuring and summarizing interaction between different types of points, one of the primary advantages of $\theta$ is that it tends to be normally distributed in its posterior distribution at the level of sub-regions (Figure S1 in the supplementary materials).  

At the end of this modeling step, we obtain a partition of each biopsy into non-overlapping sub-regions, and a measure of tumor-immune interaction on each of the sub-regions. Let $N$ be the number of biopsies, denoted $B_1, \dots, B_N$, where biopsy $B_i$ has $n_i$ observed cells $c_{i1}, \dots, c_{i{n_i}}$. For each $c_{ij}$, two pieces of information are observed: its location on the slide, $\bm{x}_{ij} = [x_{ij1}, x_{ij2}]$, and its type, $t_{ij}$, which is either ``tumor'' or ``immune.'' Next, we specify a joint model for the tumor-immune interaction across biopsies conditional on biopsy level covariates.

\subsection{Spatially structured regression model}\label{sec:regmod}

From the previous step the resulting data structure is as follows: for each biopsy $B_1, \dots, B_N$, suppose we observe on each biopsy $B_i$ a continuous outcome of interest at locations $\bm{S}_i$, i.e. $\bm{Y}_i(\bm{S}_i) = [y_i(s_{i(1)}), \dots, y_i(s_i(n_i))]^T$. Note that in our application, $y_i(s_{i(1)})$ is the  $\theta$ parameter estimated on that sub-region as described in Section \ref{sec:datastructure}. Additionally, suppose that for each biopsy, $p$ biopsy-level covariates (in our application, gene expression data) are observed, denoted by $\bm{X}_i = [x_{i1}, \dots, x_{ip}]^T$. The primary motivation for our modeling is to assess whether or not the biopsy level covariates are associated at the population level with increased heterogeneity within or between biopsies. To this end, we define the \texttt{DreameSpase} model for biopsy $i \in \{1,\ldots, N\}$ as:

% \vspace{-0.2cm}
\begin{equation}
\label{eqn:fullmodel}
\bm{Y}_i(\bm{S}_i) = (\overbrace{\boldsymbol{\alpha}^T}^{\text{Fixed effects}} \bm{X}_i) + \overbrace{\boldsymbol{\eta}_i(\bm{S}_i)}^{\text{Covariate specific random effects}} \bm{X}_i + \overbrace{\boldsymbol{\delta}_i(\bm{S}_i)}^{\text{Spatial random effect}} + \boldsymbol{\epsilon}_i.
\end{equation}
We discuss each component of the model in detail below.

\paragraph{Fixed effects} In (\ref{eqn:fullmodel}), $\boldsymbol{\alpha}^T \bm{X}_i$ is the biopsy level mean of sub-region level outcomes.
%and $\bm{1}$ is a vector of length $n_i$ where all entries are one. 
Thus, $\boldsymbol{\alpha}$ captures the variance explained by the covariates \textit{between} different biopsies and, conditional upon the covariates, can be interpreted in expectation at a population level much like the coefficients of a standard multiple regression. The object of this component of the model is to identify covariates which are systematically associated with an increase the average level of interaction in a given biopsy. 

\paragraph{Covariate specific random effects} The term $\boldsymbol{\eta}_i(\bm{S}_i)$ in (\ref{eqn:fullmodel}) is an $n_i \times p$ matrix of \textit{covariate specific spatial random effects} for each biopsy. Each column of $\boldsymbol{\eta}_i(\bm{S}_i)$ represents a spatial random effect for a specific covariate. More concretely, denoting the $j$th column of $\boldsymbol{\eta}_i$ by $\boldsymbol{\eta}_{i(j)}(\bm{S}_i)$, we assume that $\boldsymbol{\eta}_{i(j)}(\bm{S}_i)$ follows a conditionally autoregressive (CAR) process. The CAR process and other models derived from it are commonly used to analyze areal data in order to share information across adjacent sub-regions (\citealt{Orozco-Acosta22}). The form of the specific CAR process is given by:

\vspace{-0.2cm}
\begin{equation}
\label{eqn:covariateCARproc}
\boldsymbol{\eta}_{i(j)}(\bm{S}_i) \thicksim MVN(\bm{0}, \psi_j^2 \left[D_w(\bm{W}_i) - \phi \bm{W}_i\right]^{-1}).
\end{equation}
\vspace{-1.2cm}

While other options for modeling areal data exist, the CAR prior is appealing since it makes minimal assumptions about the structure of the spatial correlation beyond the dependency of regions on their neighbors as well as the fairly small number of parameters. This is particularly appealing given the non-conformability of the underlying tumor biopsies. Under this specification, $\bm{W}_i$ is an $n_i \times n_i$ symmetric matrix, where the entry in column $j$ of row $i$ (and column $i$ of row $j$) is 1 if sub-regions $i$ and $j$ are adjacent, and 0 otherwise; $D_w(\bm{W}_i)$ is a diagonal matrix where each diagonal entry is the sum of the corresponding row of $\bm{W}_i$; $\psi_j^2$ captures the variability of the value in a sub-region conditional upon the value of its neighbors; and $\phi$ captures the spatial correlation between neighbors. Note that while $\phi$ takes on values in $(-1, 1)$ and determines the correlation between adjacent sub-regions under this model specification, it is not itself \textit{equal} to the correlation between adjacent sub-regions. The distribution of this spatial random effect is thus determined by the values of $\psi_j^2$ and $\phi$.

The key parameter when considering the covariate specific spatial random effects is $\psi_j^2$. This parameter affects the outcome via the multiplication between the matrix of random effects and the covariate vector. Consider, as an extreme but illustrative example, the case where $\psi_j^2 = 0$. When this is the case, we should expect no systematic change heterogeneity \textit{within} a biopsy as the $j$th covariate changes, as each $\boldsymbol{\eta}_{i(j)}(\bm{S}_i)$ has a degenerate distribution at $\bm{0}$. However, consider $\psi_j^2 > 0$. As the magnitude of $x_{ij}$ increases, we should in turn expect to see a systematic increase in the heterogeneity of biopsy $i$. Note that this may not translate to an increase in the average \textit{level} of heterogeneity in biopsy $i$- whether or not such an increase occurs depends on the value of $\alpha_j$. These two examples, while extreme, embody the logic underlying our decision to use spike and slab priors to regularize the fittings of these parameters; details deferred to Section \ref{sec:priorspec} for now. 

\paragraph{Global CAR process} In (\ref{eqn:fullmodel}), $\boldsymbol{\delta}_i(\bm{S}_i)$ is the ``global'' spatial random effect that captures the overall spatial correlation across all biopsies. We assume that that $\boldsymbol{\delta}_i(\bm{S}_i)$ follows an analogous CAR process, defined as:

\vspace{-0.5cm}
$$
\boldsymbol{\delta}_i(\bm{S}_i) \thicksim MVN(\bm{0}, \tau^2 \left[D_w(\bm{W}_i) - \rho \bm{W}_i\right]^{-1}),
$$
where each term is defined analogously to its counterpart in (\ref{eqn:covariateCARproc}). Like the covariate specific CAR process, the distribution of this spatial random effect is thus determined by the values of $\tau^2$ and $\rho$. However, unlike the covariate specific CAR process, the $\rho$ parameter is estimated. While the model is well-defined for $\rho \in (-1, 1)$, values less than zero indicate \textit{anti}-correlation between neighbors. Because this is not a priori biologically plausible (there is no reason to suspect that adjacent tumor sub-regions should differ systematically), we assume that $\rho \in [0, 1)$, i.e. that sub-regions are on average either positively correlated or independent of their neighbors.

\paragraph{Pure error} Finally, the $\boldsymbol{\epsilon}_i \thicksim MVN(\bm{0}, \nu^2 \bm{I})$ is the pure error of the model (often referred to as the ``nugget''), i.e. the error that is not accounted for by the fixed effects, the global CAR process, or the covariate specific CAR processes.

It should be noted that the \texttt{DreameSpase} model is agnostic to how to the values on the sub-regions are generated; nothing about the downstream model depends on the usage of point processes to generate sub-regional outcome measurements. The only assumption is that the outcomes observed at the level of each sub-region are spatially correlated and (approximately) normally distributed.  

\subsection{Dual main and random effect spatial selection}\label{sec:priorspec}

\paragraph{Motivation} A challenge of working with genomic data is the dimensionality. Even when utilizing a targeted subset of available genes, it is neither likely nor expected that all genes included in the model are meaningfully related to the outcome. This necessitates some degree of selection or regularization in order to properly determine which genes are significantly associated with the outcome of interest. Performing selection yields the dual benefits of producing a more parsimonious model while also improving the accuracy of parameter estimates. To accomplish this in the context of our model, we utilize spike and slab priors for both the fixed and random effects. We focus discussion on these priors due to their importance in the model.

\paragraph{Spatial fixed effects} While there are several ways to parameterize a spike and slab prior, our method represents a standard application of \textit{stochastic search variable selection} (\cite{McCulloch93}). The prior is specified for each entry of the fixed effect vector $\alpha_j$ as follows: 

\vspace{-1cm}
$$
\alpha_j \thicksim \gamma_j N(0, \sigma^2_{slab}) + (1 - \gamma_j) N(0, \sigma^2_{spike}),
$$
$$
\gamma_j \thicksim Bernoulli(P_\gamma),
$$
where $P_\gamma$ is the prior probability of inclusion for the fixed effects and is assumed to come from a common (non-informative) $Beta(1, 1)$ prior distribution. Note that under our specification, as is often the case, $\sigma^2_{spike}$ and $\sigma^2_{slab}$ are not \textit{estimated}, but rather selected \textit{a priori}; in our application, we set $\sigma^2_{spike} = 0.03$ and $\sigma^2_{slab} = 100$. This prior construction allows not only the estimation of each $\alpha_j$ via the posterior mean, but also the probability of inclusion in the model via the posterior mean of the $\gamma_j$ parameter. Because $\gamma_j$ is an indicator, the posterior expectation corresponds to the probability that $\gamma_j = 1$, which is the probability that $\alpha_j$ is included in the model (i.e. is not penalized by the prior).

\paragraph{Spatial random effects} Unlike standard fixed effect coefficients, variance terms are constrained by definition to lie in the range $[0, \infty)$.  The challenge in adapting the spike and slab prior to the variance terms of random effects stems from the difference in domains between the variance terms $\psi_1^2, \dots, \psi_p^2$ and the fixed effect terms $\alpha_1, \dots, \alpha_p$. In theory each fixed effect term can take on any value in $\mathbb{R}$, even if excessively large or small values are implausible in practice. This is not the case when dealing with variance parameters. By definition, such parameters must lie in $[0, \infty)$. This rules out the usage of normal priors centered at zero, which put half their mass in the range $(-\infty, 0]$. Because of this asymmetry, various models have been constructed by utilizing specialized Cholesky decompositions of the covariance matrix of the random effect terms (\citealt{Joyner20}; \citealt{Ibrahim11}).
By placing penalties (in the form of priors or regularization components) on terms in the decompositions of the covariance matrix rather than directly on the terms of the covariance matrix, these models avoid dealing with both the constraints of covariance matrices as well as the domain restrictions of variance terms. As a result, such penalties can be constructed in the same manner as they might for more standard fixed effect terms. Others, such as \citealt{Scheipl12}, employ a spike and slab formulation on variance terms in the context of structural additive regression (STAR) models. However, the form and purpose of this model differs considerably from our own, as the purpose of STAR models is to provide a flexible and generalizable method for modeling the mean of some outcome vector.

We define the ``\textit{half}-spike and \textit{half}-slab'' priors for the variance terms using half-normal distributions as follows: 

\vspace*{-10pt}
$$
\psi^2_j \thicksim d_j N^+(0, \xi^2_{slab}) + (1 - d_j) N^+(0, \xi^2_{spike}),
$$
$$
d_j \thicksim Bernoulli(P_d),
$$
where $P_d \thicksim Beta(1, 1)$ is the prior probability of inclusion for all random effects. Like those of the fixed effects, the half spike is parameterized by a small variance, $\xi^2_{spike}$, while the slab is parameterized by a larger variance, $\xi^2_{slab}$. Half-normal distributions were used in place of more standard inverse gamma distributions partially due to the relatively larger amount of mass placed on extreme values by the half-normal and to achieve greater congruence between the fixed effect selection component of the model and random effect selection component of the model. This congruence makes it more straightforward to tune the selection of both the fixed and random effects in the model. The $d_j$ indicator variable also serves a similar function to the $\gamma_j$ in the fixed effect formulation.

\vspace*{15pt}
\begin{table}[!h]
    \scriptsize
    \centering
    \caption{\textbf{Full specification of \texttt{DreameSpase} model and priors.} Model specification is given for $N$ biopsies with $p$ covariates of interest. Across all terms, $i$ indexes the biopsy and $j$ indexes the covariate where relevant. $\boldsymbol{\eta}_{i(j)}$ is the $j$th column of the $n_i \times p$ matrix $\boldsymbol{\eta}_i$, the matrix of covariate specific random effects for biopsy $i$.}
    \begin{tabular}{| r c l l |}
        \hline
        &&& \\
        $\bm{Y}_i \mid \boldsymbol{\mu}_i, \nu^2$ & $\thicksim$ & $MVN(\boldsymbol{\mu}_i, \nu^2 \bm{I})$, & $i = 1, \dots, N$ \\
        $\bm{\mu}_i$ & $=$ & $\left(\boldsymbol{\alpha}^T \bm{X}_i\right) \cdot \bm{1} + \bm{\delta}_i(\bm{S}_i) + \boldsymbol{\eta}_i(\bm{S}_i) \bm{X}_i$ & $i = 1, \dots, N$ \\
        &&& \\
        $\alpha_j \mid \gamma_j, \sigma^2_{spike}, \sigma^2_{slab}$ & $\thicksim$ & $\gamma_j N(0, \sigma^2_{slab}) + (1 - \gamma_j) N(0, \sigma^2_{spike})$, & $j = 1, \dots, p$ \\
        &&& \\
        $\boldsymbol{\delta}_i(\bm{S}_i) \mid \tau^2, \rho, \bm{W}_i$ &  $\thicksim$ & $MVN(\bm{0}, \tau^2 \left[D_w(\bm{W}_i) - \rho \bm{W}_i\right]^{-1})$, & $i = 1, \dots, N$ \\
        &&& \\
        $\boldsymbol{\eta}_{i(j)}(\bm{S}_i) \mid \psi_j, \phi, \bm{W}_i$ & $\thicksim$ & $MVN(\bm{0}, \psi_j^2 \left[D_w(\bm{W}_i) - \phi \bm{W}_i\right]^{-1})$ & $i=1,\dots,N$; $j=1,\dots,p$ \\
        &&& \\
        $\psi_j \mid d_j, \xi^2_{spike}, \xi^2_{slab}$ & $\thicksim$ & $d_j N^+(0, \xi^2_{slab}) + (1 - d_j) N^+(0, \xi^2_{spike})$ & $j = 1, \dots, p$ \\
        &&& \\
        $\nu^2$, $\tau^2$ & $\thicksim$ & $Inverse~Gamma(0.001, 0.001)$ & \\
        &&& \\
        $\rho$ & $\thicksim$ & $Discrete(v_1, \dots, v_K, p_1, \dots, p_K)$ & $v_\ell \in [0, 1)$; $p_\ell \in [0,1]$;  $\sum_{\ell = 1}^K p_\ell = 1$\\
        &&& \\
        $\gamma_j \mid P_\gamma$ & $\thicksim$ & $Bernoulli(P_\gamma)$, & $j = 1, \dots, p$ \\
        &&& \\
        $d_j \mid P_d$ & $\thicksim$ & $Bernoulli(P_d)$, & $j = 1, \dots, p$ \\
        &&& \\
        $P_\gamma, P_d$ & $\thicksim$ & $Beta(1, 1)$ & \\
        &&& \\
        \hline 
    \end{tabular}
    \label{tab:modelspec}
\end{table}

\paragraph{Additional priors} The remaining priors in the model are generally chosen to be standard conjugate priors, with the exception of the $\rho$ parameter. $\rho$ is given a discrete distribution $P(\rho = v_i) = p_{v_i}$ for $i \in \{1, \dots, K\}$;  see Section \ref{sec:keyfcd} for details. As a brief point of discussion, we have chosen a non-informative $Beta(1, 1)$ prior for the prior selection probabilities for the fixed and random effects, this prior could instead be assigned uniquely to each selection probability and parameterized based on prior clinical information on the importance of each covariate in the style of \citealt{Ni19}. For full prior specifications, see Table \ref{tab:modelspec}.
\section{Posterior inference}\label{sec:postinf}

\subsection{Key full conditional distributions} \label{sec:keyfcd}

We perform posterior inference via Markov Chain Monte Carlo on the full parameter set outlined above: fixed effects $\boldsymbol{\alpha}$, corresponding selection indicators $\boldsymbol{\gamma} = [\gamma_1, \dots, \gamma_p]^T$, and overall probability of selection $P_\gamma$; random effect variance terms $\boldsymbol{\psi} = [\psi_1^2, \dots, \psi_p^2]$, corresponding selection indicators $\bm{d} = [d_1, \dots, d_p]^T$, and overall probability of selection $P_d$; the global CAR process parameters $\tau^2$ and $\rho$; and the pure error variance $\nu^2$. In this section we outline the derivations of selected full conditional distributions necessary for the implementation of the Gibbs sampling algorithm. While many of the full conditional distributions utilize conjugate or otherwise standard priors, there are several prior specifications that result in full conditional distributions unique to our model. See Table \ref{tab:modelspec} for the full model specification from which all full conditional distributions can be derived, and Section S3.3 for detailed derivations of all model parameters. Due to identifiability constraints, $\phi$ is fixed a priori at a relatively small value (0.3).

\paragraph{Spatial random effects} The full conditional of $\psi_j^2$ is induced via the prior on $\psi_j$, and requires somewhat more careful derivation. Specifying a half-normal prior on $\psi_j$ causes the full conditional distribution of $\psi_j^2$ to be that of a \textit{generalized inverse Gaussian} distribution. For a detailed derivation of this fact, see Section S3.3.2 of the supplementary materials. Thus, while the prior is specified in terms of the standard deviation $\psi_j$, in practice the sampling is performed on the variance $\psi_j^2$. However, due to the simple monotonic relationship between the two quantities, this does not pose any practical inconvenience.

%\subsubsection{Global CAR Process Correlation}\label{sec:rhoprior}

\paragraph{Global CAR process correlations} In order to improve the computational efficiency of sampling, a grid prior was placed on the correlation of the global CAR process, $\rho$. This prior is specified by a discrete set of values and corresponding probabilities, $v_1, \dots, v_K$, and $p_1, \dots, p_K$ (full conditional distributions provided in Supplementary Section S3.3.3). In practice, sampling from this distribution is accomplished using the Gumbel Max trick (\citealt{Huijben23}); see Supplementary Section S3.1  details and derivation. 

\subsection{Inferential summaries of interest}

There are two primary inferential summaries of interest: the sub-region level summaries of interaction which are summarized within and across biopsies, and the selection probabilities for the fixed effects and random effects. While we are primarily interested in the former only as the object of modeling in this paper, these summaries can be explored in many ways in their own right. As for the latter, the usage of spike and slab priors allows for the computation of posterior probabilities of selection for both the fixed and random effects $\bm{d}$ and $\bm{\Gamma}$, illustrated in panel C of Figure \ref{fig:modeloverview}. For each covariate, any combination of the corresponding fixed and random effect can be selected for inclusion in the model as explained in Section \ref{sec:priorspec}. In the context of our application: 

\begin{itemize}
    \item $P(\gamma_j = 1)$ is the probability that the $j$th fixed effect is included in the model, indicating that the corresponding gene is associated with heterogeneity \textit{between} biopsies
    \item $P(d_j = 1)$ is the probability that the $j$th random effect is included in the model, indicating that the corresponding gene is associated with heterogeneity \textit{within} biopsies
\end{itemize}
If both the fixed and random effects are selected, then a given gene is associated with both inter- and intra-biopsy heterogeneity, and thus may be of particular interest.

\section{Simulation studies}\label{sec:simulationstudy}

%\subsection{Simulation overview}\label{sec:simulationdescription}

\paragraph{Simulation overview} In order to assess the overall performance of our \texttt{DreameSpase} model we performed a series of simulation studies. In terms of design parameters, the number of biopsies $N$ was fixed at 200 (on the same order as a real data example), while the number of covariates $p$ was varied between 50, 75, and 90 to achieve ratios of $2p / N$ of 0.5, 0.75, and 0.9. We refer to the ratio $2p / N$ as the \textit{relative dimensionality} of a setting. All covariates were generated from a standard normal distribution. The number of true fixed  and random effects for each of the three settings was fixed at 3, 6, and 9, respectively. Each biopsy was simulated as a $5 \times 5$ lattice across all settings. Fixed and random effects were simulated with equal frequency to be small, medium, or large. Small, medium, and large fixed effects were simulated uniformly in the respective ranges $(0.13, 0.23)$, $(0.23, 0.33)$, and $(0.33, 0.43)$, while the corresponding ranges for random effect variance were $(0.1, 0.2)$, $(0.2, 0.3)$, and $(0.3, 0.4)$. 50 data sets were simulated for each data setting.

%\subsection{Comparative Methods}\label{sec:comparisonmethods}

\paragraph{Comparative methods}  In addition to the \texttt{DreameSpase} model, two additional models were fit to serve as performance benchmarks. The first model is a special case of the \texttt{DreameSpase} model that uses non-spatial random effects. This is equivalent to setting $\phi = 0$ and $\bm{W}_i = \mathrm{diag}(1, \ldots, 1)$ in Equation \ref{eqn:covariateCARproc}. This ignores the spatial structure of the biopsy and instead models the $\boldsymbol{\eta}_{i(j)}$ as simply clustered data. As in the true model effects were selected if their estimated posterior probability of inclusion was greater than $0.5$ in accordance with the median-probability selection model (\citealt{Barbieri04}). We refer to this model as the ``Non-Spatial \texttt{DreameSpase}'' model, abbreviated to NSDS. 

The second method was a penalization-based method that a researcher might use to answer the questions posed in this paper without developing a novel method, which we refer to as the ``analyst model.'' There were two stages of analysis for this model: one for the fixed effect component, and one for the random effect selection component. For the fixed effect selection component, a lasso regression was fit using the sub-region level outcomes as the outcome and the biopsy level covariates as the covariates. A penalization parameter was selected via three-fold cross validation, and fixed effects were considered to be selected if their parameter estimates were non-zero. To perform random effect selection, a lasso regression was fitted treating the standard deviation of each biopsy as the outcome, and the absolute value of the covariates as the regressors. In this case as well the penalization parameter was chosen via three-fold cross validation, and non-zero covariates were considered to be selected.

%\subsection{Evaluation Criteria}

\paragraph{Evaluation criteria}  The primary goal was to assess selection accuracy of the different models and methods. To do so, the true positive rate and false positive rate of selection were computed for the fixed effects as well as the random effects. The true positive rate is computed by the size of the effects, while the false positive rate is presented across all effect sizes (since null effects do not have sizes).

%\subsection{Signal to Noise Ratio}\label{sec:snr}

\paragraph{Signal to noise ratio}\label{sec:snr}  In order to properly calibrate the simulation studies to the real data and compare across differing settings, it is useful to define a signal to noise ratio. Note that unlike many models, we are interested in accurately estimating not only the fixed effect terms, but random effect terms as well. Because the two components of the model are independent of one another, we define separate signal to noise ratios for both components of the model. The fixed and random effect SNRs are defined as 

$$ SNR_{fixed} = \frac{\text{var}(\bm{X}^T \boldsymbol{\alpha})}{\text{var}(\bm{Y})}, \qquad SNR_{rand} = \frac{\sum_{j=1}^{n} \text{var}(\boldsymbol{\eta}_{j,\cdot} \bm{X})}{ n\text{var}(\bm{Y}) }. $$

See Section S2.1 for details and derivations for both quantities. The chosen simulation settings yielded simulations with fixed effect SNRs in the range $[0.37, 0.42]$, and random effect SNRs in the range $[0.38, 0.39]$.

%\subsection{Results}

\paragraph{Results}  Table \ref{tab:simTPR} shows the detailed results of the true positive rates across the effect sizes and relative dimensionality and Table \ref{tab:simFPR} shows for results of the false positive rates.

Across all settings, the false positive rate of the \texttt{DreameSpase} model never exceeded $0.02$. For medium and large effect sizes, the true positive rates for the fixed effects are uniformly $1.0$, i.e. perfect selection. For the random effects, the true positive rate lies between 0.8 and 1.0, with the median for these effect sizes being $0.98$. The model struggles substantially more with smaller effect sizes, with the lowest true positive rate obtained being 0.44 when $2p / N = 0.9$. Still, this is only marginally worse than the analyst model, which achieved the best true positive rate for this setting at $0.53$. Thus, while the model may not be sufficiently sensitive to detect all effects of this magnitude, the highly controlled false positive rate means that one can be confident that the effects we do find are true effects. Moreover, while \texttt{DreameSpase} is sometimes outperformed in selection of the smallest magnitude effects by the other methods, the differences in performance are generally quite small, while the deficiencies of the other methods across the different effect types and sizes are generally quite substantial.

The analyst model was able to select random effects with reasonable accuracy, and in fact outperforms \texttt{DreameSpase} as well as the NSDS model in the $2p / N = 0.9$ setting in terms of true positive rate, consistently achieving the best selection across several settings. The lowest true positive rate for random effect selection achieved is 0.56, (which is still the best across all three models at this setting), while the highest is 1.0. The false positive rate is also well controlled for random effects, again never exceeding 0.02. 

However, the failure to jointly model fixed effects and random effects resulted in a severely diminished true positive rate and inflated false positive rate for the fixed effects not seen when using the \texttt{DreameSpase} model. Because of the failure to jointly model the different effect types, the true positive rate for fixed effects lay between 0.4 and 0.56. Additionally, the false positive rate lay between 0.21 and 0.24, while \texttt{DreameSpase} and NSDS achieved false positive rates of less than 0.01. The only exception to this trend is the true positive rate among small effects in the $2p / N = 0.9$ setting, where the analyst model achieves a true positive rate of $0.53$ compared to 0.44 for both the \texttt{DreameSpase} as well as the NSDS model.

\begin{table}
\scriptsize
\caption{\footnotesize \textbf{True positive rate (TPR) across settings, models, and effect sizes.} Relative dimensionality is defined as $2p / N$, where $N$ is the number of biopsies and $p$ is the number of covariates. The number of covariates is doubled in the ratio to account for the estimation of both fixed effects and covariate specific random effects, thus doubling the effective dimensionality. Effect sizes refer to the magnitude of the fixed and random effects. For exact ranges, see Section \ref{sec:simulationstudy}. Bolded values indicate the best performance for a given setting and effect size. In the case of a tie, all values are bolded.}
\centering
    {\rowcolors{2}{white!90!black!50}{white!70!black!50}
    \begin{tabular}[t]{|l|l|l|l|l||l|l|l|}
    \hline
    \multicolumn{1}{|c|}{ TPR } & \multicolumn{1}{c|}{ } & \multicolumn{3}{c||}{Fixed Effects} & \multicolumn{3}{c|}{Random Effects} \\
    \cline{3-5} \cline{6-8}
    $2p / N$ & Effect Size & ``Analyst Model'' & NSDS{\textcolor{white!70!black!50}{E}} & DreameSpase & ``Analyst Model'' & NSDS{\textcolor{white!70!black!50}{E}} & DreameSpase\\
    \hline
    & Small & 0.56 & \textbf{1.00} & \textbf{1.00} & 0.94 & \textbf{0.98} & \textbf{0.98}\\
    \hline
    0.5 & Medium & 0.46 & \textbf{1.00} & \textbf{1.00} & \textbf{1.00} & \textbf{1.00} & \textbf{1.00}\\
    \hline
    & Large & 0.4 & \textbf{1.00} & \textbf{1.00} & \textbf{1.00} & \textbf{1.00} & \textbf{1.00}\\
    \hline
    \hline
    & Small & 0.48 & \textbf{0.63} & 0.61 & \textbf{0.76} & 0.26 & \textbf{0.76}\\
    \hline
    0.75 & Medium & 0.54 & \textbf{1.00} & \textbf{1.00} & \textbf{0.99} & 0.75 & 0.96\\
    \hline
    & Large & 0.51 & \textbf{1.00} & \textbf{1.00} & \textbf{1.00} & 0.93 & \textbf{1.00}\\
    \hline
    \hline
    & Small & \textbf{0.53} & 0.44 & 0.44 & \textbf{0.56} & 0.05 & 0.53\\
    \hline
    0.9 & Medium & 0.51 & \textbf{1.00} & \textbf{1.00} & \textbf{0.93} & 0.12 & 0.8\\
    \hline
    & Large & 0.49 & \textbf{1.00} & \textbf{1.00} & \textbf{0.99} & 0.2 & 0.93\\
    \hline
    \end{tabular}}
    \label{tab:simTPR}
\end{table}

For fixed effects, the performance of the NSDS model is comparable to that of \texttt{DreameSpase} across all settings in terms of both TPR and FPR. For random effects, however, the true positive rate of the NSDS model is similar to that of \texttt{DreameSpase} in the $2p / N = 0.5$ setting, but deteriorates quickly across the larger settings, at times reaching as low as 0.05. Notably, in the $2p / N = 0.9$ setting, the maximum true positive rate for random effects achieved by the NSDS model across all effect sizes is 0.2, compared to 0.93 for the \texttt{DreameSpase} model and 0.99 for the analyst model. While it achieves a false positive rate of virtually 0 in this setting, the \texttt{DreameSpase} and analyst models both achieve a modest false positive rate while also achieving a substantially higher true positive rate across all effect sizes. This indicates the importance of accounting for the spatial structure of the biopsies for the random effects. 

\paragraph{Additional simulation results} We also computed threshold free metrics, in particular Area under the curve (AUC) for all the methods and the results are shown in Table S1 of the supplementary materials. The results are broadly consistent with the ones presented above; the analyst model tends to perform well in random effect selection (normalized AUC ranging from 0.92 to 1), but struggles to a significant degree in main effect selection (normalized AUC ranging from 0.47 to 0.51).

\vspace{15pt}
\begin{table}
\footnotesize
\caption{\footnotesize \textbf{False positive rate (FPR) across settings and models.} Relative dimensionality is defined as $2p / N$, where $N$ is the number of biopsies and $p$ is the number of covariates. The number of covariates is doubled in the ratio to account for the estimation of both fixed effects and covariate specific random effects, thus doubling the effective dimensionality. Note that false positive rates are determined solely by the selection results for the simulated null covariates, and thus the FPR does not vary when broken down by effect size. Bolded values indicate the best performance for a given setting and effect size. In the case of a tie, all values are bolded.}
\centering
{\rowcolors{3}{white!90!black!50}{white!70!black!50}
\begin{tabular}[t]{|l|l|l|l||l|l|l|}
\hline
\multicolumn{1}{|c|}{ FPR } & \multicolumn{3}{c||}{Fixed Effects} & \multicolumn{3}{c|}{Random Effects} \\
\cline{2-4} \cline{5-7}
$2p / N$ & ``Analyst Model'' & NSDS & DreameSpase & ``Analyst Model'' & NSDS & DreameSpase\\
\hline
0.5 &  0.24 & \textbf{< 0.01} & \textbf{< 0.01} & \textbf{< 0.01} & \textbf{< 0.01} & 0.02\\
\hline
0.75 & 0.22 & \textbf{< 0.01} & \textbf{< 0.01} & 0.01 & \textbf{< 0.01} & 0.01\\
\hline
0.9 & 0.21 & \textbf{< 0.01} & \textbf{< 0.01} & 0.02 & \textbf{< 0.01} & 0.01\\
\hline
\end{tabular}}
\label{tab:simFPR}
\end{table}

\section{Tumor microenvironment in melanoma}\label{sec:application}

\subsection{Background}

\paragraph{Melanoma and immunotherapy} Melanoma is a particularly interesting type of cancer given its high immunogenicity caused by its high mutational load (\citealt{Marzagalli19}). This characteristic has implications not only for patient prognosis, but also susceptibility to treatment via immunotherapies (\citealt{Kang20}; \citealt{Simiczyjew20}). While remarkably effective in certain patients, immunotherapy is by no means a silver bullet in the treatment of melanoma. Ipilimumab, the first CTLA-4 inhibitor approved by the FDA for treatment of melanoma patients, has demonstrated remarkable efficacy in the treatment of even advanced stage melanoma, both alone and in concert with other treatments. However, the objective response rate across these different settings is far from 100\% across all patients (\citealt{Chesney18}). % ; \citealt{Margolin12}; \citealt{Hodi10}). 
Further complicating this picture is the fact that the composition of TME has also been shown to be associated with response to immunotherapies (\citealt{Falcone20}). 
The biological understanding of the interplay between immunotherapy and the composition of the TME has been bolstered recently by the proliferation of single cell imaging technologies (\citealt{Xiao22}).
Still, more work is required to synthesize our understandings of immunological mechanisms in melanoma and its relationship to the spatial composition of the TME. To this end, we investigate the relationship between spatial composition of the TME (using pathology imaging data) and immunological mechanisms (using genomic data), both at the level of tumor-immune interactions as well as the heterogeneity of the interactions within and across tumors.

\paragraph{Description of pathology imaging and genomic data} Our motivating imaging dataset comes from the The Cancer Imaging Archive (\citealt{tcia13}), a companion consortium to the The Cancer Genome Atlas project (TCGA). 
TCGA is an initiative that makes available high quality genomics and imaging data on samples from a diverse range of cancer types (\citealt{Weinstein13}). This multi-platform data includes, most notably for our purposes, high definition images of the biopsies and gene expression data from patient samples. 
Our pathology imaging data set consisted of whole slide imaging data for 335 biopsies from the Skin Cutaneous Melanoma (SKCM), with the number of cells per biopsy ranging from 814 to 187,521. The cells of each biopsy were processed and classified by a machine learning model as either tumor or immune according to the morphology of each cell. Biopsies were subsequently partitioned into non-overlapping sub-regions, and an estimate of interaction was computed for each sub-region. For more details on the cell classification algorithm and tumor partitioning algorithm, see \cite{Osher23}. 

In order to determine the association between immunological mechanisms and spatial association of tumor cells and immune cells, we used gene expression data from genes associated with different groups of immune cells. These immune groups include 
``marker'' genes linked to B cells, macrophages, monocytes, neutrophils, natural killer cells, plasma cells, and T cells. These immune cell groups regulate many important oncogenic processes such as angiogenesis, metastasis, and the response to drugs or immunotherapy
(\citealt{Nirmal18}).

Using the sub-regional interaction parameter estimates described in Section \ref{sec:datastructure} as the outcome, the \texttt{DreameSpase} model was fit using gene expression data for the previously described sets of genes downloaded from the UCSC Xena browser (\citealt{Goldman20}). In total, there were 11,202 sub-regions on which interaction was estimated across 335 biopsies, with the median number of sub-regions per biopsy being 29 (IQR 26). After pre-processing, there were 122 genes included in our final analysis and additional pre-processing details are available in Section S3.2 of the supplementary materials; see Algorithm S1 for the exact algorithm used.

The \texttt{DreameSpase} model was run for 300,000 total iterations, 150,000 of which were treated as burn-in. The resulting samples were thinned by 10, resulting in 15,000 posterior samples. Global convergence for the posterior samples was assessed via the Geweke diagnostic on the likelihood ($p > 0.05$); see Figure S2 of the supplementary materials for trace plots of log-likelihood and selected parameters. The analysis took approximately 8 hours on an M1 Macbook Pro with 8 gigabytes of memory.

\subsection{Results}

\begin{figure}[H]
    \centering
    \includegraphics[width=13cm]{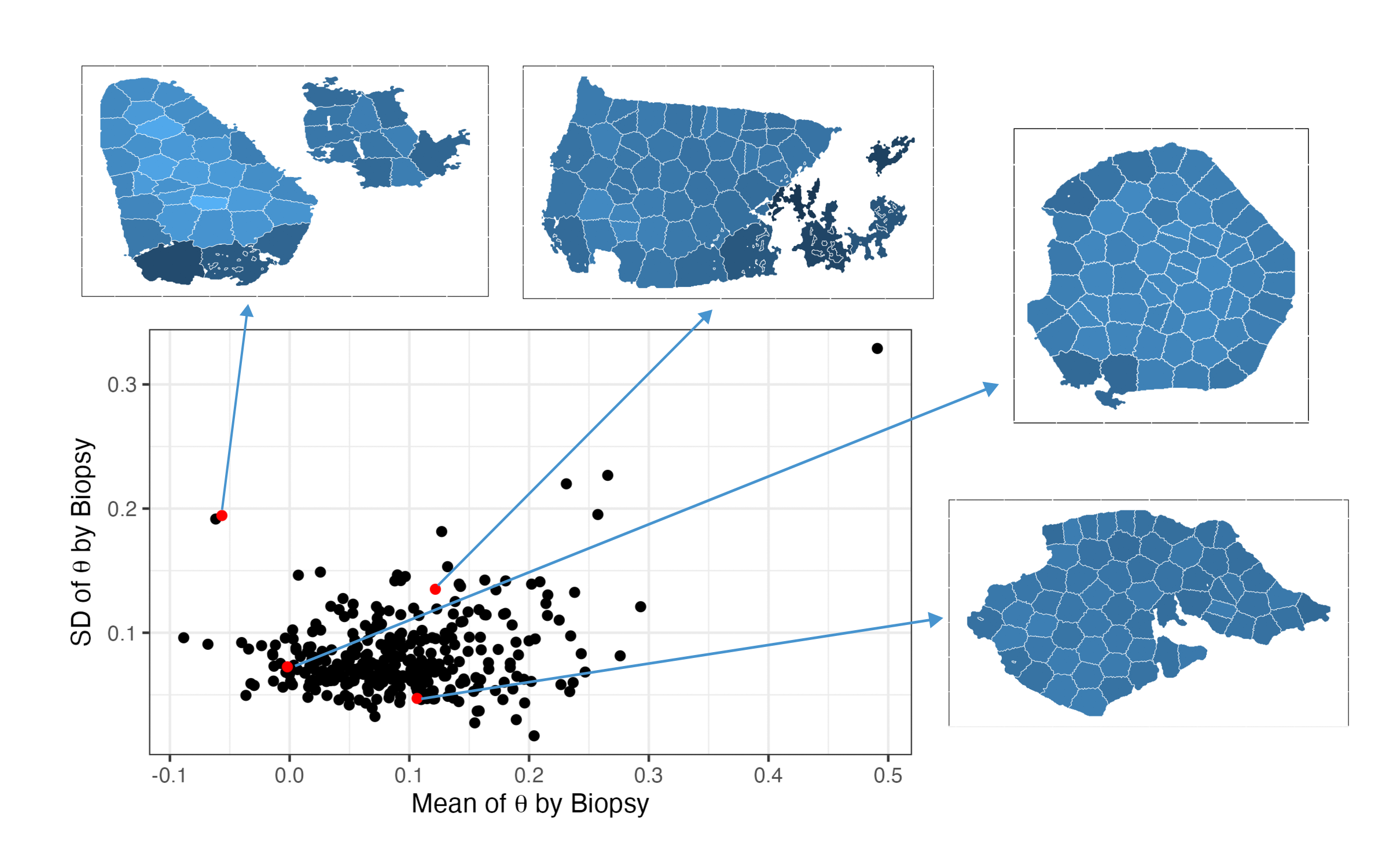}
    \caption{\textbf{Distribution of spatial interactions within and across biopsies.} Standard deviation of  cellular spatial interaction parameter ($\theta$) plotted against mean $\theta$ parameter at the biopsy level. Also shown are selected biopsy level plots of $\theta$ by sub-region (darker shades imply higher interaction) .}
    \label{fig:interactiondistribution}
\end{figure}

\paragraph{Overview} Figure \ref{fig:interactiondistribution} illustrates the spatial heterogeneity of the tumor-immune interaction both within biopsies (captured by the standard deviation of sub-regional $\theta$ estimates) and between them (captured by the mean of the $\theta$ estimates). Biopsy level means ranged from -0.09 to 0.49, while biopsy level standard deviations ranged from 0.02 to 0.32. The absolute value of the coefficients of variation also ranged considerably, the minimum being 0.08 and the maximum being 437. This illustrates the degree to which the average level of interaction within a biopsy can vary separately from the level of heterogeneity within a biopsy.

\begin{figure}
    \centering
    \includegraphics[width=17cm]{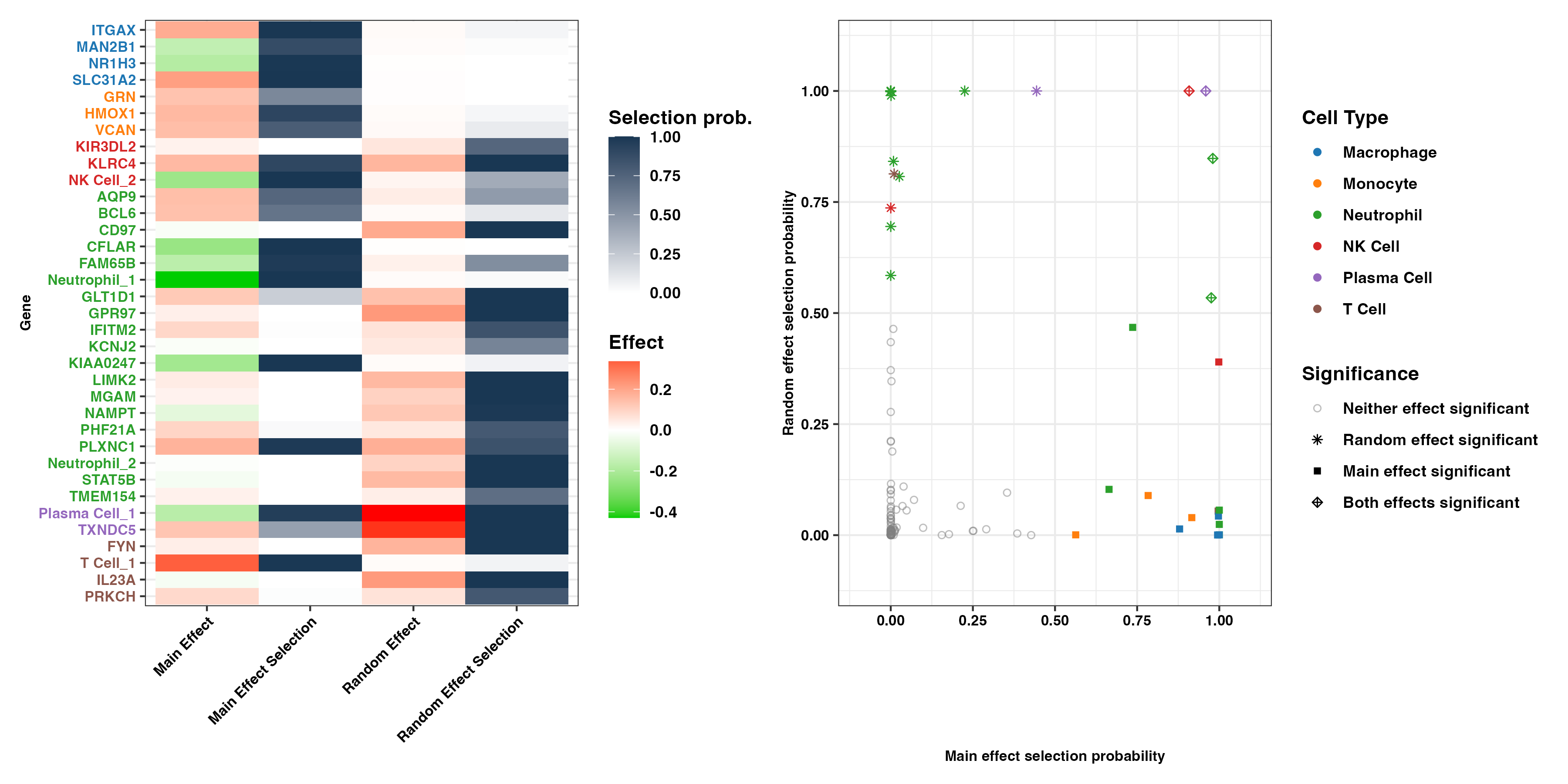}
    \caption{\textbf{Left panel: heatmap of selection probabilities and effect magnitude for fixed and random effects for selected genes.} Posterior mean and selection probabilities for main effects and random effects. Gene names are colored according to immune class membership. \textbf{Right panel: Selection results from gene expression analysis.} Posterior probability of random effect selection plotted against posterior probability of fixed effect selection. Color indicates class of immune cells the gene belongs to; shape indicates whether or not the effect was selected using the median-probability selection model described in Section \ref{sec:simulationstudy}. }
    \label{fig:selectionresults}
\end{figure}

The left panel of Figure \ref{fig:selectionresults} shows the relative magnitudes of the selected fixed and random effects, while the right panel shows the results of selection for fixed effects as well as random effects by immune cell sub-class. As described in Section \ref{sec:simulationstudy}, fixed effects and random effects were considered selection if their posterior probability of selection was greater than 0.5. Ultimately, there were 35 covariates for which the fixed effect, random effect, or both were determined to be significant- see Table S3 of the supplementary materials for the full list along with immune cell group and effect selection.
These genes came from 6 of the 7 total immune cell subtypes: macrophages, monocytes, neutrophils, natural killer cells, plasma cells, and T cells. The key results from the groups with selected genes are detailed in the following paragraphs.

\paragraph{Neutrophils} The immune cell group with the largest number of effects selected was the Neutrophil group. Neutrophils play a complex role in the tumor microenvironment, having been observed to have both pro- and anti-tumorigenic impacts depending on context (\citealt{Powell16}). Notably, one of the genes for which both the fixed effect and random effect was selected was PLXNC1, from the Neutrophil gene set. Increased expression of PLXNC1 was associated with increased tumor-immune interaction between biopsies. This is consistent with previous biological research that has suggested that PLXNC1 may act as a tumor suppressor, stymying progression and metastasis (\citealt{scott09}). The combined gene covariate Neutrophil\_1 consisted of the genes S100A8 and S100A9, and was negatively associated with interaction. This is highly consistent with previous results that indicate that expression of one or both of these genes is inversely associated with overall survival in melanoma patients (\citealt{Pour21}). We thus find evidence in our results for the temperamental nature of neutrophil presence in the tumor microenvironment.

\paragraph{T cells} While often treated as a monolith, the T cell family consists of many sub-types which play vastly differing roles in the TME. For example, while CD8+ T cells are widely regarded as a key part of the anti-tumorigenic immune response, regulatory T cells can actually have a pro-tumorigenic impact by suppressing immune response (\citealt{Xie21}).%, \citealt{Li20}). 
The combined gene covariate T\_1 consisted of 72 genes, the full list of which is available in Table S2 of the supplementary materials, and was positively associated with interaction. While interpretation is intrinsically more difficult due to the number of genes being combined, this set most notably contained the genes CD8A (often referred to as CD8) and CD3E (often referred to as CD3). The presence of one or both of these genes has been used as a proxy for the presence of cytotoxic lymphocytes in the context of multiplex imaging (\citealt{Qin22}). It is thus unsurprising that their expression at the biopsy level would be associated with the presence of such cytotoxic cells, and thus increased interaction between immune cells and tumor cells. Finally, among the genes where the random effects were selected was IL23A. IL23A is a pro-inflammatory cytokine that has been shown to be associated with invasiveness in melanoma (\citealt{klein15}). %, \citealt{croxford12}). 
While a number of different phenotypes could correspond to increased heterogeneity within a biopsy, it is intuitive that such heterogeneity would be associated with increased inflammation and invasiveness into the surrounding tissue.

\paragraph{Monocytes} The role of monocytes in the tumor microenvironment is complex, primarily due to the fact that monocytes can differentiate into both macrophages and dendritic cells, each of which has different implications for patient prognosis. While tumor-associated macrophages are generally associated with poor patient prognosis, tumor-associated dendritic cells are generally associated with positive prognosis (\citealt{Ugel21}). Amongst the selected genes in the monocyte group, two stood out based on their prior established importance in melanoma: HMOX1 and VCAN. HMOX1 has been shown to be genetically linked to risk of melanoma (\citealt{Okamoto06}). Further, downregulating HMOX1 has been shown to be associated with increased interaction between natural killer cells and the tumor (\citealt{Furfaro20}). This is somewhat incongruous with our result, since the estimated fixed effect term for HMOX1 is positive; however, since we are essentially adjusting for natural killer cell presence by proxy (i.e. expression of genes related to natural killer cells), such an association is not a priori biologically impossible. It has also been established that VCAN is overexpressed in melanoma relative to comparable healthy cells and that VCAN is associated with increased proliferation and in certain melanoma lineages (\citealt{Touab02}; \citealt{Hern11}).

\section{Discussion and limitations}\label{sec:conclusionsandlimitations}

In this paper, we propose a method called \texttt{DreameSpase} for integrative analyses of digital pathology imaging and genomics data. \texttt{DreameSpase}  provides a novel spatially structured regression framework for joint modeling of high resolution spatial data from non-conformable spaces (i.e. biopsies) and associated (biopsy-level) covariates (i.e. gene expression data). This is  accomplished by positing a
structured regression model with covariate-specific spatial random effects which are modelled using CAR-based spatial processes. Furthermore, we induce sparsity by generalizing the traditional spike and slab priors used for fixed effect selection to the selection of spatial random that captures the intra-space heterogeneity.  We demonstrate via simulation studies that the model reliably detects inter-space heterogeneity via the fixed effect selection mechanism, and intra-space heterogeneity via the random effect selection mechanism. Finally, we demonstrate the utility of \texttt{DreameSpase} in the analysis of whole slide pathology imaging data by applying it to a set of images of melanoma biopsies and associated transcriptomics data.

Advances in scale and extent of imaging technologies has led to proliferation of quantitative imaging data across many diseases, especially cancer. This has allowed a systematic collation of high-resolution spatial datasets -- at scale -- across many patients. These include newer technologies such as spatial multiplexed tissue imaging that generates high-resolution images to study the associations of cellular spatial relationships with tumor growth, metastasis, drug resistance, and patient survival (\citealt{van_dam_multiplex_2022}). Our model engenders joint modeling of such non-conformable spatial data, as well as assessment of inter and intra-patient heterogeneity and its association with patient level covariates.

There remain areas of improvement for the model. In our formulation, the selection of the fixed effects is modeled independently of the selection of the random effects. This may be a reasonable assumption, but there may be circumstances in which it may be desirable to allow for the modeling of a broader class of relationships between the selection of the two types of effects. Another shortcoming is the formulation of the random effect selection. The random effect selection suffers from a ``piranha problem'' whereby if too many null random effects are included in the model, it can be challenging for the model to properly select random effects. This is because each random effect, even if unselected, must have positive variance and thus account for some small part of the variance of the outcome. In practice, the estimated values of these random effects tend to be quite small individually, but large enough in aggregate that if too many are included, the ability of the model to select true random effects may suffer. In addition to these areas of improvement, there are also natural potential areas of extension. While we have limited our inquiry to genomic data, the model could naturally be extended in the future to multi-omic data (e.g. proteomic, metabolomic data). While other types of data could simply be included as covariates, it would perhaps be desirable to introduce group or more structured priors in such a setting (\citealt{Zhang14}). Such priors could be used to select broad types of covariates (e.g. proteomics vs. genomics) but also potential pathways and families of genes. We leave these tasks for future exploration. 

\paragraph{Software and data availability} We have developed a general purpose software package that is available as a public Github repository. The package includes an efficient implementation of the Gibbs sampler algorithm discussed in this paper as well as methods to simulate data from the model. The sampling algorithm is implemented in C++ for improved efficiency. Data used in the analysis are provided in a separate Github repository with scripts to reproduce the analysis, tables and figures. This repository is modular in nature so that individual tables and figures can be reproduced as necessary. 
\setlength{\bibsep}{0pt}


\begin{thebibliography}{}

\bibitem[Abousamra et~al., 2022]{Abousamra22}
Abousamra, S., Gupta, R., Hou, L., Batiste, R., Zhao, T., Shankar, A., Rao, A.,
  Chen, C., Samaras, D., Kurc, T., and Saltz, J. (2022).
\newblock Deep learning-based mapping of tumor infiltrating lymphocytes in
  whole slide images of 23 types of cancer.
\newblock {\em Frontiers in Oncology}, 11:806603.

\bibitem[Barbieri and Berger, 2004]{Barbieri04}
Barbieri, M.~M. and Berger, J.~O. (2004).
\newblock Optimal predictive model selection.
\newblock {\em The Annals of Statistics}, 32(3).

\bibitem[Baxi et~al., 2022]{Baxi22}
Baxi, V., Edwards, R., Montalto, M., and Saha, S. (2022).
\newblock Digital pathology and artificial intelligence in translational
  medicine and clinical practice.
\newblock {\em Modern Pathology}, 35(1):23–32.

\bibitem[Bowman, 2014]{Bowman14}
Bowman, F.~D. (2014).
\newblock Brain imaging analysis.
\newblock {\em Annual Review of Statistics and Its Application}, 1(1):61–85.

\bibitem[Chervoneva et~al., 2021]{Chervoneva21}
Chervoneva, I., Peck, A.~R., Yi, M., Freydin, B., and Rui, H. (2021).
\newblock Quantification of spatial tumor heterogeneity in immunohistochemistry
  staining images.
\newblock {\em Bioinformatics}, 37(10):1452–1460.

\bibitem[Chesney et~al., 2018]{Chesney18}
Chesney, J., Puzanov, I., Collichio, F., Singh, P., Milhem, M.~M., Glaspy, J.,
  Hamid, O., Ross, M., Friedlander, P., Garbe, C., Logan, T.~F., Hauschild, A.,
  Lebbé, C., Chen, L., Kim, J.~J., Gansert, J., Andtbacka, R.~H., and Kaufman,
  H.~L. (2018).
\newblock Randomized, open-label phase ii study evaluating the efficacy and
  safety of talimogene laherparepvec in combination with ipilimumab versus
  ipilimumab alone in patients with advanced, unresectable melanoma.
\newblock {\em Journal of Clinical Oncology}, 36(17):1658–1667.

\bibitem[Clark et~al., 2013]{tcia13}
Clark, K., Vendt, B., Smith, K., Freymann, J., Kirby, J., Koppel, P., Moore,
  S., Phillips, S., Maffitt, D., Pringle, M., Tarbox, L., and Prior, F. (2013).
\newblock The cancer imaging archive (tcia): Maintaining and operating a public
  information repository.
\newblock {\em Journal of Digital Imaging}, 26(6):1045–1057.

\bibitem[Falcone et~al., 2020]{Falcone20}
Falcone, I., Conciatori, F., Bazzichetto, C., Ferretti, G., Cognetti, F.,
  Ciuffreda, L., and Milella, M. (2020).
\newblock Tumor microenvironment: Implications in melanoma resistance to
  targeted therapy and immunotherapy.
\newblock {\em Cancers}, 12(10):2870.

\bibitem[Furfaro et~al., 2020]{Furfaro20}
Furfaro, A.~L., Ottonello, S., Loi, G., Cossu, I., Piras, S., Spagnolo, F.,
  Queirolo, P., Marinari, U.~M., Moretta, L., Pronzato, M.~A., Mingari, M.~C.,
  Pietra, G., and Nitti, M. (2020).
\newblock Ho‐1 downregulation favors brafv600 melanoma cell death induced by
  vemurafenib/plx4032 and increases nk recognition.
\newblock {\em Int. J. Cancer}.

\bibitem[George and McCulloch, 1993]{McCulloch93}
George, E.~I. and McCulloch, R.~E. (1993).
\newblock Variable selection via gibbs sampling.
\newblock {\em Journal of the American Statistical Association},
  88(423):881–889.

\bibitem[Goldman et~al., 2020]{Goldman20}
Goldman, M.~J., Craft, B., Hastie, M., Repe{\v{c}}ka, K., McDade, F., Kamath,
  A., Banerjee, A., Luo, Y., Rogers, D., Brooks, A.~N., et~al. (2020).
\newblock Visualizing and interpreting cancer genomics data via the xena
  platform.
\newblock {\em Nature biotechnology}, 38(6):675--678.

\bibitem[Hanahan, 2022]{Hanahan22}
Hanahan, D. (2022).
\newblock Hallmarks of cancer: New dimensions.
\newblock {\em Cancer Discovery}, 12(1):31–46.

\bibitem[Heindl et~al., 2015]{Heindl15}
Heindl, A., Nawaz, S., and Yuan, Y. (2015).
\newblock Mapping spatial heterogeneity in the tumor microenvironment: a new
  era for digital pathology.
\newblock {\em Laboratory Investigation}, 95(4):377–384.

\bibitem[Hernández, 2011]{Hern11}
Hernández, D. (2011).
\newblock V3 versican isoform alters the behavior of human melanoma cells by
  interfering with cd44/erbb-dependent signaling.
\newblock {\em Journal of Biological Chemistry}, 286(2).

\bibitem[Huijben et~al., 2023]{Huijben23}
Huijben, I. A.~M., Kool, W., Paulus, M.~B., and Van~Sloun, R. J.~G. (2023).
\newblock A review of the gumbel-max trick and its extensions for discrete
  stochasticity in machine learning.
\newblock {\em IEEE Transactions on Pattern Analysis and Machine Intelligence},
  45(2).

\bibitem[Högmander and Särkkä, 1999]{Hogmander99}
Högmander, H. and Särkkä, A. (1999).
\newblock Multitype spatial point patterns with hierarchical interactions.
\newblock {\em Biometrics}, 55(4):1051–1058.

\bibitem[Ibrahim et~al., 2011]{Ibrahim11}
Ibrahim, J.~G., Zhu, H., Garcia, R.~I., and Guo, R. (2011).
\newblock Fixed and random effects selection in mixed effects models.
\newblock {\em Biometrics}, 67(2):495–503.

\bibitem[Joyner et~al., 2020]{Joyner20}
Joyner, C.~N., McMahan, C.~S., Tebbs, J.~M., and Bilder, C.~R. (2020).
\newblock From mixed effects modeling to spike and slab variable selection: A
  bayesian regression model for group testing data.
\newblock {\em Biometrics}, 76(3):913–923.

\bibitem[Kang et~al., 2020]{Kang20}
Kang, K., Xie, F., Mao, J., Bai, Y., and Wang, X. (2020).
\newblock Significance of tumor mutation burden in immune infiltration and
  prognosis in cutaneous melanoma.
\newblock {\em Frontiers in Oncology}, 10:573141.

\bibitem[Klein et~al., 2015]{klein15}
Klein, A., Schwartz, H., Sagi-Assif, O., Meshel, T., Izraely, S., Ben~Menachem,
  S., Bengaiev, R., Ben-Shmuel, A., Nahmias, C., Couraud, P.-O., et~al. (2015).
\newblock Astrocytes facilitate melanoma brain metastasis via secretion of
  il-23.
\newblock {\em The Journal of pathology}, 236(1):116--127.

\bibitem[Krishnan et~al., 2022]{Krishnan22}
Krishnan, S.~N., Mohammed, S., Frankel, T.~L., and Rao, A. (2022).
\newblock Gawrdenmap: a quantitative framework to study the local variation in
  cell–cell interactions in pancreatic disease subtypes.
\newblock {\em Scientific Reports}, 12(1):3708.

\bibitem[Li et~al., 2019a]{Li19}
Li, Q., Wang, X., Liang, F., and Xiao, G. (2019a).
\newblock A bayesian mark interaction model for analysis of tumor pathology
  images.
\newblock {\em The Annals of Applied Statistics}, 13(3).

\bibitem[Li et~al., 2019b]{LiPotts19}
Li, Q., Wang, X., Liang, F., Yi, F., Xie, Y., Gazdar, A., and Xiao, G. (2019b).
\newblock A bayesian hidden potts mixture model for analyzing lung cancer
  pathology images.
\newblock {\em Biostatistics}, 20(4):565–581.

\bibitem[Li et~al., 2021]{Li21}
Li, X., Yang, Y., Huang, Q., Deng, Y., Guo, F., Wang, G., and Liu, M. (2021).
\newblock Crosstalk between the tumor microenvironment and cancer cells: A
  promising predictive biomarker for immune checkpoint inhibitors.
\newblock {\em Frontiers in Cell and Developmental Biology}, 9:738373.

\bibitem[MacNab, 2022]{MacNab22}
MacNab, Y.~C. (2022).
\newblock Bayesian disease mapping: Past, present, and future.
\newblock {\em Spatial Statistics}, 50:100593.

\bibitem[Marzagalli et~al., 2019]{Marzagalli19}
Marzagalli, M., Ebelt, N.~D., and Manuel, E.~R. (2019).
\newblock Unraveling the crosstalk between melanoma and immune cells in the
  tumor microenvironment.
\newblock {\em Seminars in Cancer Biology}, 59:236–250.

\bibitem[Ni et~al., 2019]{Ni19}
Ni, Y., Stingo, F.~C., and Baladandayuthapani, V. (2019).
\newblock Bayesian graphical regression.
\newblock {\em Journal of the American Statistical Association},
  114(525):184–197.

\bibitem[Nirmal et~al., 2018]{Nirmal18}
Nirmal, A.~J., Regan, T., Shih, B.~B., Hume, D.~A., Sims, A.~H., and Freeman,
  T.~C. (2018).
\newblock Immune cell gene signatures for profiling the microenvironment of
  solid tumors.
\newblock {\em Cancer Immunology Research}, 6(11):1388–1400.

\bibitem[Okamoto et~al., 2006]{Okamoto06}
Okamoto, I., Kr, J., Mannhalter, C., Wagner, O., and Pehamberger, H. (2006).
\newblock A microsatellite polymorphism in the heme oxygenase‐1 gene promoter
  is associated with risk for melanoma.
\newblock {\em International Journal of Cancer}.

\bibitem[Orozco-Acosta et~al., 2022]{Orozco-Acosta22}
Orozco-Acosta, E., Adin, A., and Ugarte, M.~D. (2022).
\newblock Big problems in spatio-temporal disease mapping: methods and
  software.
\newblock {\em Computer Methods and Programs in Biomedicine},
  231(arXiv:2201.08323).
\newblock arXiv:2201.08323 [stat].

\bibitem[Osher et~al., 2023]{Osher23}
Osher, N., Kang, J., Krishnan, S., Rao, A., and Baladandayuthapani, V. (2023).
\newblock Spartin: a bayesian method for the quantiﬁcation and
  characterization of cell type interactions in spatial pathology data.
\newblock {\em Frontiers in Genetics}.

\bibitem[Pour et~al., 2021]{Pour21}
Pour, S.~R., de~Coa{\~n}a, Y.~P., Demorentin, X.~M., Melief, J., Thimma, M.,
  Wolodarski, M., Gomez-Cabrero, D., Hansson, J., Kiessling, R., and Tegner, J.
  (2021).
\newblock Predicting anti-pd-1 responders in malignant melanoma from the
  frequency of s100a9+ monocytes in the blood.
\newblock {\em Journal for immunotherapy of cancer}, 9(5).

\bibitem[Powell and Huttenlocher, 2016]{Powell16}
Powell, D.~R. and Huttenlocher, A. (2016).
\newblock Neutrophils in the tumor microenvironment.
\newblock {\em Trends in Immunology}, 37(1):41–52.

\bibitem[Qin et~al., 2022]{Qin22}
Qin, A., Lima, F., Bell, S., Kalemkerian, G.~P., Schneider, B.~J., Ramnath, N.,
  Lew, M., Krishnan, S., Mohammed, S., Rao, A., and Frankel, T.~L. (2022).
\newblock Cellular engagement and interaction in the tumor microenvironment
  predict non-response to pd-1/pd-l1 inhibitors in metastatic non-small cell
  lung cancer.
\newblock {\em Scientific Reports}, 12(1):9054.

\bibitem[Sadeghi~Rad et~al., 2021]{Sadeghi21}
Sadeghi~Rad, H., Monkman, J., Warkiani, M.~E., Ladwa, R., O’Byrne, K.,
  Rezaei, N., and Kulasinghe, A. (2021).
\newblock Understanding the tumor microenvironment for effective immunotherapy.
\newblock {\em Medicinal Research Reviews}, 41(3):1474–1498.

\bibitem[Saltz et~al., 2018]{Saltz18}
Saltz, J., Gupta, R., Hou, L., Kurc, T., Singh, P., Nguyen, V., Samaras, D.,
  Shroyer, K.~R., Zhao, T., Batiste, R., et~al. (2018).
\newblock Spatial organization and molecular correlation of tumor-infiltrating
  lymphocytes using deep learning on pathology images.
\newblock {\em Cell reports}, 23(1):181--193.

\bibitem[Scheipl et~al., 2012]{Scheipl12}
Scheipl, F., Fahrmeir, L., and Kneib, T. (2012).
\newblock Spike-and-slab priors for function selection in structured additive
  regression models.
\newblock {\em Journal of the American Statistical Association},
  107(500):1518–1532.

\bibitem[Scott et~al., 2009]{scott09}
Scott, G.~A., McClelland, L.~A., Fricke, A.~F., and Fender, A. (2009).
\newblock Plexin c1, a receptor for semaphorin 7a, inactivates cofilin and is a
  potential tumor suppressor for melanoma progression.
\newblock {\em Journal of Investigative Dermatology}, 129(4):954--963.

\bibitem[Simiczyjew et~al., 2020]{Simiczyjew20}
Simiczyjew, A., Dratkiewicz, E., Mazurkiewicz, J., Zi\c{e}tek, M., Matkowski,
  R., and Nowak, D. (2020).
\newblock The influence of tumor microenvironment on immune escape of melanoma.
\newblock {\em International Journal of Molecular Sciences}, 21(21):8359.

\bibitem[Sun, 2016]{Sun16}
Sun, Y. (2016).
\newblock Tumor microenvironment and cancer therapy resistance.
\newblock {\em Cancer Letters}, 380(1):205–215.

\bibitem[Touab, 2002]{Touab02}
Touab, M. (2002).
\newblock Versican is differentially expressed in human melanoma and may play a
  role in tumor development.
\newblock {\em The American Journal of Pathology}, 160(2).

\bibitem[Ugel et~al., 2021]{Ugel21}
Ugel, S., Canè, S., De~Sanctis, F., and Bronte, V. (2021).
\newblock Monocytes in the tumor microenvironment.
\newblock {\em Annual Review of Pathology: Mechanisms of Disease},
  16(1):93–122.

\bibitem[Van~Dam et~al., 2022]{van_dam_multiplex_2022}
Van~Dam, S., Baars, M. J.~D., and Vercoulen, Y. (2022).
\newblock Multiplex {Tissue} {Imaging}: {Spatial} {Revelations} in the {Tumor}
  {Microenvironment}.
\newblock {\em Cancers}, 14(13):3170.

\bibitem[Vu et~al., 2022]{Vu22}
Vu, T., Wrobel, J., Bitler, B.~G., Schenk, E.~L., Jordan, K.~R., and Ghosh, D.
  (2022).
\newblock Spf: A spatial and functional data analytic approach to cell imaging
  data.
\newblock {\em PLOS Computational Biology}, 18(6):e1009486.

\bibitem[Weinstein et~al., 2013]{Weinstein13}
Weinstein, J.~N., Collisson, E.~A., Mills, G.~B., Shaw, K. R.~M., Ozenberger,
  B.~A., Ellrott, K., Shmulevich, I., Sander, C., and Stuart, J.~M. (2013).
\newblock The cancer genome atlas pan-cancer analysis project.
\newblock {\em Nature Genetics}, 45(10):1113–1120.

\bibitem[Xiao et~al., 2022]{Xiao22}
Xiao, X., Guo, Q., Cui, C., Lin, Y., Zhang, L., Ding, X., Li, Q., Wang, M.,
  Yang, W., Kong, Y., and Yu, R. (2022).
\newblock Multiplexed imaging mass cytometry reveals distinct tumor-immune
  microenvironments linked to immunotherapy responses in melanoma.
\newblock {\em Communications Medicine}, 2(1):131.

\bibitem[Xie et~al., 2021]{Xie21}
Xie, Q., Ding, J., and Chen, Y. (2021).
\newblock Role of cd8+ t lymphocyte cells: Interplay with stromal cells in
  tumor microenvironment.
\newblock {\em Acta Pharmaceutica Sinica B}, 11(6):1365–1378.

\bibitem[Yang et~al., 2017]{Yang17}
Yang, H., Baladandayuthapani, V., and Morris, J.~S. (2017).
\newblock Quantile functional regression using quantlets.
\newblock {\em arXiv:1711.00031 [stat]}.
\newblock arXiv: 1711.00031.

\bibitem[Zhang et~al., 2014]{Zhang14}
Zhang, L., Baladandayuthapani, V., Mallick, B.~K., Manyam, G.~C., Thompson,
  P.~A., Bondy, M.~L., and Do, K.-A. (2014).
\newblock Bayesian hierarchical structured variable selection methods with
  application to molecular inversion probe studies in breast cancer.
\newblock {\em Journal of the Royal Statistical Society Series C: Applied
  Statistics}, 63(4):595–620.

\bibitem[Zoghi et~al., 2023]{zoghi23chap1}
Zoghi, S., Masoumi, F., and Rezaei, N. (2023).
\newblock The immune system.
\newblock In {\em Clinical Immunology}, pages 1--46. Elsevier.

\end{thebibliography}
\end{document}

\maketitle
\tableofcontents

\renewcommand{\thesection}{S\arabic{section}}
\renewcommand{\thefigure}{S\arabic{figure}}
\renewcommand{\thetable}{S\arabic{table}}
\renewcommand{\theequation}{S\arabic{equation}}
\renewcommand{\thealgorithm}{S\arabic{algorithm}}
\setstretch{1.2}

\section{Pre-Processing}

\subsection{H\&E Data}

335 high definition images were obtained from The Cancer Genome Atlas Genomic Data Commons Data Portal. 20 images were used to train a cell classification model. A total of 1,250 cells were annotated by a pathologist. These annotated cells were subsequently used to train a random forest model to classify the types of cells in each image as Tumor, Immune, Macrophage, or Other based on morphology characteristics of the cell nuclei. The classifier to achieved an accuracy in the range of 87\%-91\%. In addition to the type of each cell, the x- and y- coordinates of the centroid of each cell were determined. This process therefore yielded three pieces of information for each cell in the biopsy: the type, the x-coordinate, and the y-coordinate. For more details on the classifier and pre-processing, see \citealt{Osher23}. % \redacted{redacted}.

\subsection{Hierarchical Strauss Model Fitting}

Model fitting for the Hierarchical Strauss Model (HSM) is based on the Pseudolikelihood function, which is itself based on the Papangelou conditional intensity function. Let $\mathbf{\Theta} = [\beta_1, \beta_2, \theta]^T$ be the parameters of the HSM, $\boldmath{x} = [\boldmath{x}_1^T, \boldmath{x_2}^T]^T$ be the observed points, and let the density $f(\boldmath{x}; \boldsymbol{\Theta})$ be the likelihood evaluated on a given point pattern $\boldmath{x}$ for a set of parameters $\boldsymbol{\Theta}$. Then the Papangelou conditional intensity at a point $u$ is defined by:

\begin{equation}
\label{eqn:CL}
\lambda(u|\boldsymbol{\Theta}, \boldmath{x}) = 
\begin{cases}
\frac{f(\boldmath{x} \cup \{u\}; \boldsymbol{\Theta})}{f(\boldmath{x}; \boldsymbol{\Theta})} & u \not\in \boldmath{x} \\
\frac{f(\boldmath{x}; \boldsymbol{\Theta})}{f(\boldmath{x} - \{u\}; \boldsymbol{\Theta})} & u \in \boldmath{x}
\end{cases}
\end{equation}

The pseudolikelihood (PL) is in turn defined based on the Papangelou conditional intensity. Because the PL involves an integral, a quadrature based approximation is often used in place of the PL function (and in turn the log-PL function). A $\boldmath{u}$ is selected, with corresponding weights $\boldmath{w}$. The pseudolikelihood function can then be defined as follows:

\begin{equation}
\label{eqn:PLapprox}
PL(\boldsymbol{\Theta} | \boldmath{x}) \approx
\prod_{x_i \in \boldmath{x}} \lambda(x_i | \boldsymbol{\Theta}, \boldmath{x}) \exp\left(-\sum_{u_j \in \boldmath{u}} \lambda(u_j | \boldsymbol{\Theta}, \boldmath{x}) w_j\right)
\end{equation}

Yielding a log-PL function of:

\begin{equation}
\label{eqn:logPLapprox}
logPL(\boldsymbol{\Theta} | \boldmath{x}) \approx
\sum\limits_{x_i \in \boldmath{x}} \log\left(\lambda(x_i | \boldsymbol{\Theta}, \boldmath{x})\right) -\sum_{u_j \in \boldmath{u}} \lambda(u_j | \boldsymbol{\Theta}, \boldmath{x}) w_j
\end{equation}

The log-PL function is used as the target log-likelihood function, and traditional sampling methods can be used. We utilized the STAN statistical software to perform the posterior sampling of the $\boldsymbol{\Theta}$ parameter.

\subsection{Posterior Distributions of $\theta$}

\begin{figure}[H]
    \centering
    \includegraphics[width=17cm]{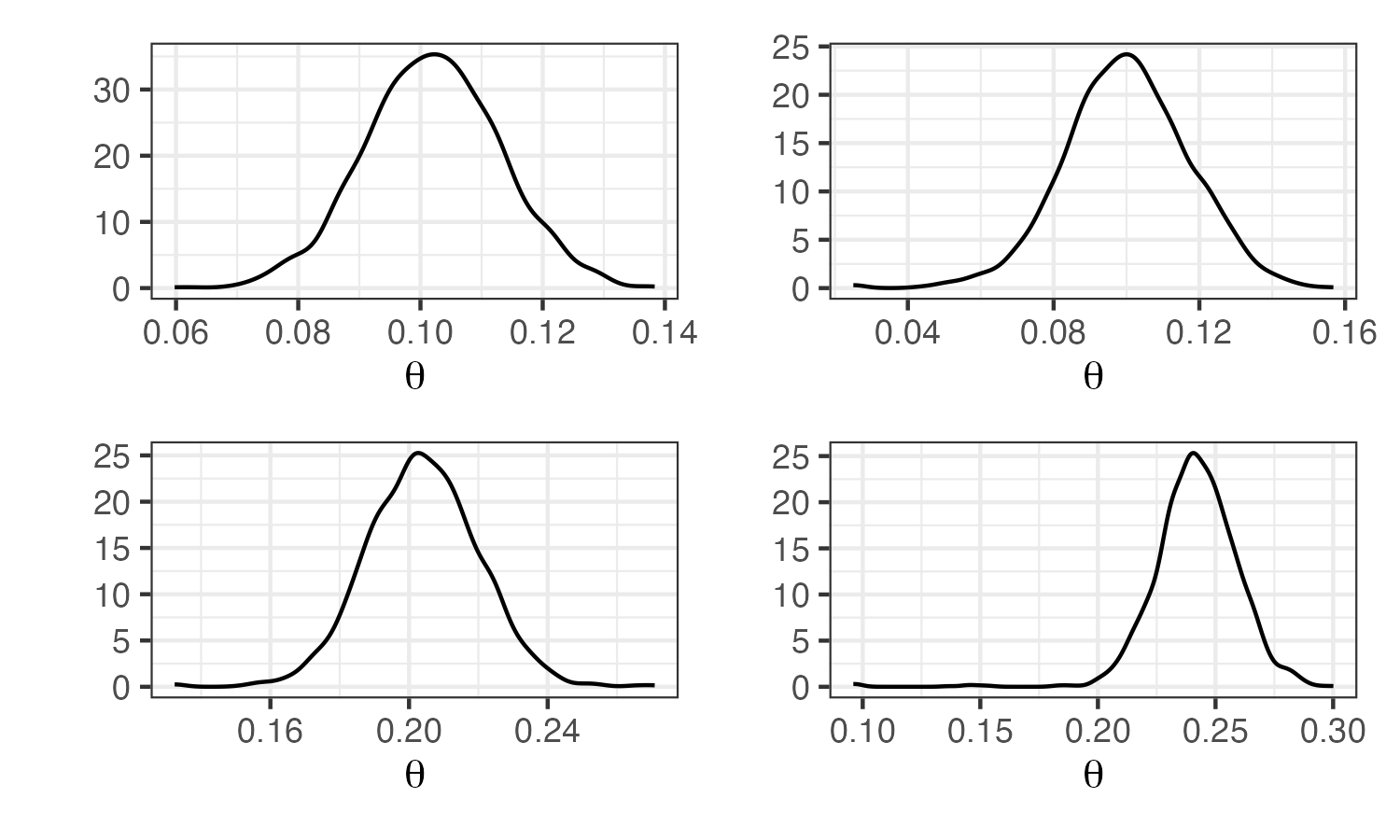}
    \caption{\textbf{Posterior distributions of $\theta$.} Example posterior distributions of $\theta$ parameter for various sub-regions of biopsies in the data set.}
    \label{fig:thetaposteriors}
\end{figure}

Figure \ref{fig:thetaposteriors} shows examples of posterior distributions for the $\theta$ parameter for various sub-regions of various biopsies. These posteriors tend to be relatively Gaussian.

\section{Simulation}

\subsection{Signal to Noise Ratio}

\subsubsection{Fixed Effect Signal to Noise Ratio}

Let us assume that the covariates follow some distribution with covariance matrix $\boldsymbol{\Sigma}_x$ and mean $\boldsymbol{\mu}_x = [\mu_1, \dots, \mu_p]^T$. Then denoting the marginal variance of the outcome $\boldmath{Y}$ across all biopsies by $\text{var}(\boldmath{Y})$:

\vspace{-0.4cm}
\begin{equation}
\label{eqn:snrmain}
SNR_{fixed} = \frac{\text{var}(\boldmath{X}^T \boldsymbol{\alpha})}{\text{var}(\boldmath{Y})}
\end{equation}

This can be thought of as an $R^2$ value, in that it represents the proportion of variance explained by the fixed effects. When $\boldsymbol{\alpha}$ is treated as a fixed constant, by basic properties of covariance we have \(
\text{var}(\boldmath{X}^T \boldsymbol{\alpha}) = \boldsymbol{\alpha}^T\boldsymbol{\Sigma}_x \boldsymbol{\alpha}
\). Because this value depends only on the joint distribution of the covariates, the values of the fixed effects, and the marginal distribution of the outcome, it is common across all biopsies.

\subsubsection{Random Effect Signal to Noise Ratio}

The signal to noise ratio for the random effects is defined analogously to those of the fixed effects. However, the signal to noise ratio must be considered at the level of a given biopsy rather than across all biopsies. This is because of the assumptions of the CAR model: biopsies with adjacency structures such that sub-regions tend to have fewer neighbors will have greater variance associated with the spatial random effects, and vice versa for adjacency structures such that sub-regions tend to have more neighbors. Note that this also means that within a biopsy, the level of variance attributable to the spatial random effect for a given sub-region will depend on the number of neighbors that sub-region has. 

Dropping patient index $i$ for ease of notation, let $\boldsymbol{\eta}_{j,\cdot}$ denote the $j$th row of $\boldsymbol{\eta}_i(\boldmath{S}_i)$. Then define the mean SNR for a biopsy with adjacency matrix $W$ and $n$ sub-regions by:

\begin{equation}
\label{eqn:snrrand}
SNR_{rand} = \frac{\sum_{j=1}^{n} \text{var}(\boldsymbol{\eta}_{j,\cdot} \boldmath{X})}{ n\text{var}(\boldmath{Y}) }
\end{equation}

In other words, the signal to noise for the random effect component for a given biopsy is the ratio of the average variance accounted for by the random effect across the sub-regions. Note that because the variance of the outcome for a given biopsy may be larger than that of the marginal variance of the outcome, this term cannot be interpreted as a proportion of variance explained in the same manner as the fixed effect SNR, since it can be larger than one. 

 Because both $\boldsymbol{\eta}_{j,\cdot}$ and $\boldmath{X}$ are being treated as random quantities, the term $\text{var}(\boldsymbol{\eta}_{j,\cdot} \boldmath{X})$ requires modestly more thought. Because the entries of $\boldsymbol{\eta}_{j,\cdot}$ are independent, it is easiest to think of this as \(
\text{var}(\sum_{k=1}^p \eta_{jk} X_{k})
\), which by the law of total variance and the independence of the entries of $\boldsymbol{\eta}_{j,\cdot}$ is equal to

\(
\text{var}\left(E\left[\sum_{k=1}^p  \eta_{jk} X_{k} | \boldsymbol{\eta}_{j,\cdot}\right]\right) + 
E\left[\text{var}\left(\sum_{k=1}^p  \eta_{jk} X_{k} | \boldsymbol{\eta}_{j,\cdot}\right)\right]
\)

\(
= \text{var}\left(\sum_{k=1}^p \eta_{jk} \mu_k \right) + E\left[\boldsymbol{\eta}_{j,\cdot} \boldsymbol{\Sigma}_x \boldsymbol{\eta}_{j,\cdot}^T \right]
\)

\(
= \sum_{k=1}^p \mu_k^2 \text{var}(\eta_{jk}) + \sum_{k=1}^p E[\eta_{jk}^2]\sigma^2_k
\)

Where $\sigma^2_k$ is the $k$th diagonal entry of $\boldsymbol{\Sigma}_x$. Because $E[\boldsymbol{\eta}_{j,\cdot}] = \boldmath{0}$, this reduces to

\(
= \sum_{k=1}^p \text{var}(\eta_{jk})(\mu_k^2 + \sigma^2_k) 
\)

Here again $\text{var}(\eta_{jk})$ refers to the variance of the $j$th covariate specific CAR random effect for the $k$th covariate, which is the $j$th entry of $\psi_k^2 [D_w(W) - \phi W]^{-1}$; again this cannot be written more simply.

Finally, note that by definition $\text{var}(\epsilon_i) = \nu^2$. 

\subsection{Additional Simulation Results}

Define $AUC_{p}$ as the area under the ROC curve limited to false positive rate $p$. This means that rather than lying between 0 and 1 like standard $AUC$, $AUC_{p}$ will lie between 0 and $p$. Table \ref{tab:simAUC} shows the $AUC_{0.1}$ and $AUC_{0.2}$ for the three models. In order to compute $AUC_{p}$, the TPR/FPR pair for $FPR = p$ was estimated by linearly interpolating the observed pair with the largest FPR less than $p$ and the observed pair with the smallest FPR greater than $p$. For example, if we observe an FPR/TPR pair of $0.19, 0.49$ and $0.21, 0.51$, we would estimate that $TPR = 0.52$ when $FPR = 0.2$. The area under the curve is computed empirically, i.e. treating the observed pairs as a step function and integrating it.

\begin{table}
    \caption{\textbf{$\boldmath{AUC_{0.1}}$ and $\boldmath{AUC_{0.2}}$ for simulations}. $AUC_{0.1}$ refers to the $AUC$ computed such that the maximal FPR considered is $0.1$, with $AUC_{0.2}$ defined analogously. Thus, $AUC_{0.1} \in [0, 0.1]$, and $AUC_{0.2} \in [0, 0.2]$. For the purposes of the tables below, both have been multiplied by 10 and 5, respectively, to put them in the same range of AUC (between 0.5 and 1). The ``analyst model'' performs quite well for random effect selection, but notably less well for fixed effect selection. The \textit{DreameSpase} model performs marginally worse than the ``analyst model'' in Random Effect selection in the latter two settings, but considerably better in the fixed effect setting. }
    \centering
    {\rowcolors{2}{white!90!black!50}{white!70!black!50}
    \begin{tabular}[t]{|l|l|l|l|l|l|l|}
    \hline
    \multicolumn{1}{|c|}{ $\mathbf{AUC_{0.1}}$ } & \multicolumn{3}{c|}{Main Effects} & \multicolumn{3}{c|}{Random Effects} \\
    \cline{2-4} \cline{5-7}
    \hline
    $\frac{2p}{n}$ & ``Analyst Model'' & NSDS & \textit{DreameSpase} & ``Analyst Model'' & NSDS & 
    \textit{DreameSpase}\\
    \hline
    0.5 & 0.47 & \textbf{1} & \textbf{1} & \textbf{0.99} & \textbf{0.99} & \textbf{0.99}\\
    \hline
    0.75 & 0.51 & \textbf{1} & \textbf{1} & \textbf{0.97} & 0.82 & 0.94\\
    \hline
    0.9 & 0.51 & \textbf{1} & \textbf{1} & \textbf{0.92} & 0.65 & 0.81\\
    \hline
    \end{tabular}}
    {\rowcolors{2}{white!90!black!50}{white!70!black!50}
    \begin{tabular}[t]{|l|l|l|l|l|l|l|}
    \hline
    \multicolumn{1}{|c|}{ $\mathbf{AUC_{0.2}}$ } & \multicolumn{3}{c|}{Main Effects} & \multicolumn{3}{c|}{Random Effects} \\
    \cline{2-4} \cline{5-7}
    \hline
    $\frac{2p}{n}$ & ``Analyst Model'' & NSDS & \textit{DreameSpase} & ``Analyst Model'' & NSDS & 
    \textit{DreameSpase}\\
    \hline
    0.5 & 0.47 & \textbf{1} & \textbf{1} & \textbf{1} & 0.99 & 0.99\\
    \hline
    0.75 & 0.51 & \textbf{1} & \textbf{1} & \textbf{0.98} & 0.84 & 0.96\\
    \hline
    0.9 & 0.51 & \textbf{1} & \textbf{1} & \textbf{0.95} & 0.7 & 0.84\\
    \hline
    \end{tabular}}
    \label{tab:simAUC}
\end{table}

\section{Application}

\subsection{Gumbel Max Trick Derivation}\label{sec:gumbelmax}

Suppose $\gamma_1,...,\gamma_k$ are log non-normalized probabilities of a categorical distribution; we will treat these values as fixed. Let $G(\mu)$ denote the Gumbel distribution with scale parameter 1 and location parameter $\mu$. Denote the CDF of each distribution by $F_\mu(x) = \exp(-\exp(-(x - \mu)))$ and the PDF by $f_\mu(x) = \exp(-(x - \mu) - \exp(-(x - \mu)))$. Let $g_1,...,g_k$ be samples with $g_j \thicksim G(\gamma_j)$. Consider the probability that $g_j$ is the largest among all $g_1,...,g_k$. Conditional upon the value of $g_j$, the probability that $g_j$ is the largest is simply

\begin{fleqn}
\begin{equation}
\label{eqn:gmderiv1}
\begin{alignedat}{2}
P(g_1,...,g_{j-1}, g_{j+1},...,g_k < g_j | g_j)  &= \prod_{i \neq j} P(g_i < g_j) \\ 
 &=  \prod_{i \neq j} F_{\gamma_i}(g_j) \\
 & = \prod_{i\neq j} \exp(-\exp(-(g_j - \gamma_{i}))) = \exp(-\exp(-g_j)\sum_{i\neq j}\exp(\gamma_i))
\end{alignedat}
\end{equation}
\end{fleqn}

Thus,

\begin{fleqn}
\begin{equation}
\label{eqn:gmderiv2}
\begin{alignedat}{2}
P(g_1,...,g_{j-1}, g_{j+1},...,g_k < g_j) &= E[I(g_1,...,g_{j-1}, g_{j+1},...,g_k < g_j)] \\
&= E_{g_j}[E[I(g_1,...,g_{j-1}, g_{j+1},...,g_k < g_j) | g_j]] \\
&= E_{g_j}[\exp(-\exp(-g_j)\sum_{i\neq j}\exp(\gamma_i))] \\
&= \int_{-\infty}^\infty \exp(-\exp(-g_j)\sum_{i\neq j}\exp(\gamma_i)) \exp(-(g_j - \gamma_j) - \exp(-(g_j - \gamma_j))) dg_j \\ 
&= \int_{-\infty}^\infty \exp(-(g_j - \gamma_j) -\exp(-g_j)\sum_{i=1}^k\exp(\gamma_i))  dg_j \\
&= \frac{\exp(\gamma_j)}{\sum_{i=1}^k\exp(\gamma_i)}\exp(-\exp(-g_j)\sum_{i=1}^k\exp(\gamma_i)) |_{-\infty}^\infty = \frac{\exp(\gamma_j)}{\sum_{i=1}^k\exp(\gamma_i)} \\
\end{alignedat}
\end{equation}
\end{fleqn}

Which is precisely the marginal probability of drawing from category $j$ in the categorical distribution.

\subsection{Gene Expression Pre-Processing}

The pre-processing of the gene expression data used in the application analysis proceeded as follows. Within the seven groups of genes, the pairwise correlation was computed between all pairs of genes. Genes were then gathered into sets such that for all genes in a given group, there was at least one other gene in that set for which the pairwise correlation was at least 0.8. Note that such sets would consist of a single gene if a given gene was not highly correlated with any other genes. The expression of all genes within a given group were then averaged to yield a single value, which was ultimately used in the analysis. 

Equivalently, this process can be thought of as forming a graph within each group where each gene represents a vertex, and there is an edge between two vertices (genes) if their pairwise correlation is greater than 0.8. Each subgroup represents a disjoint sub-graph of the resulting graph. This process is described in the Algorithm $\ref{alg:clustering}$. This process resulted in 9 sets with more than one gene. The sets and the genes they contain are shown in Table \ref{tab:genecombos}.

\begin{algorithm}
\caption{Gene Expression Pre-Processing}\label{alg:clustering}
\begin{algorithmic}
\For{$G_i$ in $\{$Gene group 1, $\dots$, Gene group 7$\}$}
    \For{$k = 1\dots N_{G_i}$} \Comment{$N_{G_i}$ is the number of genes in group $G_i$}
        \If{$g_{ik}$ is not currently in a set}
            \State create a new set, $C_{ik} = \{g_{ik}\}$
        \EndIf
        
        \For{$j = i + 1 \dots N_{G_i}$}
            \If{$Cor(g_{ij}, g_{ik}) > 0.8$}
                \State merge sets of $g_{ij}, g_{ik}$
            \EndIf
        \EndFor        
    \EndFor
\EndFor

\For{$C_{\ell m}$ in $C_{11}, \dots, C_{1N_{C_1}}, \dots, C_{71}, \dots, C_{7N_{C_7}}$} \Comment $N_{C_i}$ is the number of sets for group $i$ 
    \State compute $E_{\ell m} = \frac{1}{|C_{\ell m}|} \sum_{t=1}^{|C_{\ell m}|} g_t$
\EndFor

\Return $E_{11}, \dots, E_{1N_{C_1}}, \dots, E_{71}, \dots, E_{7N_{C_7}}$
\end{algorithmic}
\end{algorithm}

\begin{table}[]
    \centering
    \caption{\textbf{Gene sets and component genes.} Each gene set consists of two or more gene, with all genes in the group having a correlation between their expression of at least 0.8. The average expression of each set is included in the model as a single covariate.}
    \begin{tabular}{|c|p{100mm}|}
        \hline
        Gene set & Genes \\
        \hline
         B\_1 & BANK1; HLA-DOB; CD72; TLR10; CD19; TCL1A; MS4A1; STAP1; BTLA; 
         CR2; FCRL2; CD180; VPREB3; FCRL1; FCRL3; FAM129C; FCRL5; CD79A; CCR6; LY9; CD37;  KIAA0125; PNOC; PAX5; POU2F2; S1PR4; BLK  \\
         \hline
         Macrophage\_1 & CECR1; SLAMF8; IFI30; CCR1; CD163; ITGB2; C1QB; C3AR1; FCER1G; TYROBP; TNFAIP2; SLC15A3; CD74; CLEC7A; NCKAP1L; SPI1; CYBB; VSIG4; HK3; IGSF6; MSR1; LILRB4; CD300A; TLR8; MNDA; FCGR1B; FPR3; FCGR1A; CD4; MYO1F; CYTH4; CD86; LAIR1; LAPTM5; ADAMDEC1; CMKLR1; MS4A7; TNFRSF1B; MS4A4A; CTSS; AOAH; ITGAM; CSF1R; C1QA; C5AR1; ATP8B4; CCRL2; SLCO2B1 \\ 
         \hline
         Neutrophil\_1 & FPR1; CSF3R; LILRA2; NCF4 \\
         \hline
         T\_1 & GIMAP4; CD2; ARHGAP9; CD48; RASSF5; CD52; ARHGAP25; TBC1D10C; NLRC3; C1orf162; SP140; GPR18; HCST; RHOH; GZMK; CORO1A; ITGAL; GIMAP7; TRAF3IP3; EVI2B; DOCK2; IL10RA; LCP1; CD27; FAM26F; DOCK8; CD3G; GIMAP2; NCF1B; FLI1; CXCR6; SH2D1A; PVRIG; CYTIP; TRAT1; CD3E; GIMAP6; CD96; CD3D; CRTAM; CCL19; BIN2; PARVG; TARP; KLRB1; CCR7; CD6; UBASH3A; PSTPIP1; IL7R; GPR171; APBB1IP; AMICA1; BTK; PTPRCAP; ITK; SLA; GIMAP5; RCSD1; SASH3; TNFRSF9; CD28; HVCN1; CXCL9; LY86; RGS18; DPEP2; SIRPG; CD8A; ICOS; GAB3; GMFG \\
         \hline
         NK\_1 & KLRC3; KLRC2 \\
         \hline
         Monocyte\_1 & PILRA; LILRB2; HCK; LILRB3; LST1; AIF1; LILRA6; CD300LF; FGR; CD14; C10orf54; SLC7A7; NFAM1; PRAM1; LRRC25 \\
         \hline
         Neutrophil\_2 & S100A9; S100A8 \\
         \hline
         NK\_2 & TBX21; KIR2DL4; PRF1; SAMD3; KLRD1 \\
         \hline
         Plasma\_1 & TNFRSF17; IGJ \\
         \hline
    \end{tabular}
    \label{tab:genecombos}
\end{table}

\subsection{Full Conditional Distributions}

This section outlines the derivations of the full conditional distributions for all parameters in the model. The derivation is divided into three parts: one for the main effects and associated terms, one for the random effects and associated terms, and one for the error terms.

\subsubsection{Main Effect Group}

\paragraph{Main Effects}

We begin by defining the following quantities:

$\boldmath{Y}_{i,\alpha}^* = \boldmath{Y}_i - \boldsymbol{\eta}_i \boldmath{X}_i - \boldsymbol{\delta}_i(\boldmath{S}_i)$, $\overline{Y}_{i,\alpha}^* = \frac{1}{n_i} (\boldmath{1}^T \boldmath{Y}_{i,\alpha}^*)$ (i.e. the mean of each $\boldmath{Y}_{i,\alpha}^*$), and $\boldmath{\overline{Y}}_\alpha^* = [\overline{Y}_{1,\alpha}^*, \dots, \overline{Y}_{N,\alpha}^*]^T$. Note that $\pi(\overline{Y}_{i,\alpha}^* \mid \cdot)\thicksim N(\boldsymbol{\alpha}^T \boldmath{X}_i, \frac{\nu^2}{n_i^2})$.

Further, defining the $N \times p$ matrix $\boldmath{X}$ such that $[\boldmath{X}]_{ij} = X_{ij}$, i.e. the $j$th covariate of the $i$th patient, then trivially we have $\pi(\boldsymbol{\alpha} \mid \cdot) \propto \pi(\boldmath{\overline{Y}}_\alpha^* \mid \boldsymbol{\alpha}, \nu^2) \pi(\boldsymbol{\alpha} \mid \gamma_1, \dots, \gamma_p, \sigma^2_{spike}, \sigma^2_{slab})$. Denoting $\boldsymbol{\Gamma} = diag(\gamma_1 \cdot \sigma_{slab}^2 + (1 - \gamma_1) \cdot \sigma_{spike}^2, \dots, \gamma_p \cdot \sigma_{slab}^2 + (1 - \gamma_p) \cdot \sigma_{spike}^2)$, i.e. the prior covariance of $\boldsymbol{\alpha}$ conditional on $\gamma_1, \dots, \gamma_p$, and denoting $\boldmath{D} = diag(\frac{\nu^2}{n_1^2}, \dots, \frac{\nu^2}{n_p^2})$ the full conditional for $\boldsymbol{\alpha}$ is 

$$
\pi(\boldsymbol{\alpha} \mid \cdot) = MVN( (\boldsymbol{\Gamma}^{-1} + \nu^{-2}\boldmath{X}^T\boldmath{D}^{-1}\boldmath{X})^{-1} \boldmath{X}^T \boldmath{D}^{-1} \boldmath{\overline{Y}}_\alpha^*, (\boldsymbol{\Gamma}^{-1} + \nu^{-2}\boldmath{X}^T\boldmath{D}^{-1}\boldmath{X})^{-1})
$$

This is simply the standard full conditional distribution for the main effects of a linear regression with an informative prior and heteroschedastic error variance.

\paragraph{Main Effect Selection Indicators}

For each main effect selection indicator $\gamma_j$, $j = 1, \dots, p$, by definition $\gamma_j$ takes on the value either 0 or 1. Thus, given the prior $P(\gamma = 1) = p_{\gamma,1}$ and $P(\gamma = 0) = p_{\gamma,0}$, it follows that $P(\gamma_j = k \mid \cdot) \propto \pi(\alpha_j \mid \gamma, \cdot) p_{\gamma,k}$ for $k = 0,1$. Defining $\sigma^2_k = k \sigma^2_{slab} + (1 - k) \sigma^2_{spike}$, this implies that the full conditional for $\gamma_j$ is given by:

$$
P(\gamma = k \mid \cdot) = \frac{\normdenszero{\sigma_k}{\alpha_j}p_{\gamma,k}}{\normdenszero{\sigma_{spike}}{\alpha_j}p_{\gamma,0} + \normdenszero{\sigma_{slab}}{\alpha_j}p_{\gamma,1}}
$$

\paragraph{Main Effect Selection Probability}\label{sec:maineffselectionprob}

The final term in the main effect group is the main effect selection probability term, $P_\gamma$. This term captures the proportion of main effects selected. The combination of the uninformative beta prior in $P_\gamma$ and the binary $\gamma_1,\dots,\gamma_p$, the full conditional of $P_\gamma$ trivially reduces to that of a standard beta-binomial model:

$$
    \pi(P_\gamma \mid \cdot) = Beta\left(1 + \sum_{j=1}^p \gamma_j, 1 + \left[p - \sum_{j=1}^p \gamma_j\right]\right)
$$

\subsubsection{Random Effect Group}

\paragraph{Spatial Random Effects}\label{sec:spatialrandomeffects}

The full conditional of $\psi_j^2$ is induced via the prior on $\psi_j$, and requires somewhat more careful derivation. We begin by noting that $\pi(\psi_j \mid \cdot) \propto \left[\prod_{i=1}^N \pi(\boldsymbol{\eta}_{i(j)}(\boldmath{S}_i) \mid \psi_j, \cdot) \right] \pi(\psi_j \mid d_j, \xi^2_{spike}, \xi^2_{slab})$. More precisely, defining $\xi^2_k = k \xi^2_{slab} + (1 - k) \xi^2_{spike}$ this yields the full conditional kernel of:

\begin{fleqn}
\begin{equation}
\label{eqn:psifc1}
\begin{alignedat}{2}
    \pi(\psi_j \mid d_j = k, \cdot) \propto & \prod\limits_{i=1}^N \Biggl[ \text{det}\left( \frac{1}{\psi_j^2} \left[ D_w(\boldmath{W}_i) - \phi \boldmath{W}_i \right] \right)^{-1/2}  \\
    & \exp\left(-\frac{1}{2}\left\{{\boldsymbol{\eta}_{i(j)}(\boldmath{S}_i)}^T \left[ 
    \frac{1}{\psi_j^2} \left[ D_w(\boldmath{W}_i) - \phi \boldmath{W}_i \right]\right]^{-1} \boldsymbol{\eta}_{i(j)}(\boldmath{S}_i)\right\}\right)
    \Biggr] \\
    & \normdenszeroprop{\xi_k}{\psi_j} \\
\end{alignedat}
\end{equation}
\end{fleqn}
\newcommand{\eqnpsifcone}{7 }

Let us denote $S = \sum_{i=1}^N n_i$, i.e. the total number of sub-regions across all biopsies. Using properties of the determinant and standard algebraic manipulation, equation \eqnpsifcone can be simplified to 

\begin{fleqn}
\begin{equation}
\label{eqn:psifc2}
\begin{alignedat}{2}
    \pi(\psi_j \mid d_j = k, \cdot) \propto 
    (\psi_j)^{-S} 
    \exp\left(-\frac{1}{2}\left[\sum_{i=1}^N 
    \boldsymbol{\eta}_{i(j)}(\boldmath{S}_i)^T  
    \left[ D_w(\boldmath{W}_i) - \phi \boldmath{W}_i \right] 
    \boldsymbol{\eta}_{i(j)}(\boldmath{S}_i)
    \right]\frac{1}{\psi_j^2} - 
    \frac{1}{2\xi_k^2}\psi_j^2
    \right)
\end{alignedat}
\end{equation}
\end{fleqn}
\newcommand{\eqnpsifctwo}{8 }

While it appears unusual at first glance, the right hand side of equation \eqnpsifctwo is the kernel of a \textit{Generalized Inverse Gaussian} distribution, parameterized, the kernel of which is given by: 

$$
f(x \mid a, b, c) \propto x^{c - 1} \exp\left(-\frac{1}{2}\left(ax + \frac{b}{x}\right)\right)
$$

However, this is with respect to $\psi_j^2$, while the density itself is of $\psi_j$. Fortunately, because the variance $\psi_j^2$ is a monotonic transformation of the standard deviation $\psi_j$, it follows from this that the distribution of $\psi_j^2$ is given by

\begin{fleqn}
\begin{equation}
\label{eqn:psifc3}
\begin{alignedat}{2}
    \pi(\psi_j^2 \mid d_j = k, \cdot) &\propto
    ((\psi_j^2)^{1/2})^{-S} 
    \left| \frac{1}{2}(\psi_j^2)^{-1/2} \right| \\
    &\exp\left(-\frac{1}{2}\left[\sum_{i=1}^N 
    \boldsymbol{\eta}_{i(j)}(\boldmath{S}_i)^T  
    \left[ D_w(\boldmath{W}_i) - \phi \boldmath{W}_i \right] 
    \boldsymbol{\eta}_{i(j)}(\boldmath{S}_i)
    \right]\frac{1}{((\psi_j^2)^{1/2})^2} - 
    \frac{1}{2\xi_k^2}((\psi_j^2)^{1/2})^2
    \right) \\ 
    &=(\psi_j^2)^{-(S + 1)/2}
    \exp\left(-\frac{1}{2}\left\{\left[\sum_{i=1}^N 
    \boldsymbol{\eta}_{i(j)}(\boldmath{S}_i)^T  
    \left[ D_w(\boldmath{W}_i) - \phi \boldmath{W}_i \right] 
    \boldsymbol{\eta}_{i(j)}(\boldmath{S}_i)
    \right]\frac{1}{\psi_j^2} + 
    \frac{1}{\xi_k^2}\psi_j^2\right\}
    \right)
\end{alignedat}
\end{equation}
\end{fleqn}
\newcommand{\eqnpsifcthree}{10 }

Where the term $\left| \frac{1}{2}(\psi_j^2)^{-1/2} \right|$ is the Jacobian of the transformation. It follows from this that

$$
\pi(\psi_j^2 \mid d_j = k, \cdot) = 
GIG(a = \frac{1}{\xi_k^2}, 
    b = \sum_{i=1}^N 
\boldsymbol{\eta}_{i(j)}(\boldmath{S}_i)^T  
\left[ D_w(\boldmath{W}_i) - \phi \boldmath{W}_i \right] 
\boldsymbol{\eta}_{i(j)}(\boldmath{S}_i),
c = \frac{-S}{2} + \frac{1}{2})
$$

Thus, while the prior is specified in terms of the standard deviation $\psi_j$, in practice the sampling is performed on the variance $\psi_j^2$. However, due to the simple monotonic relationship between the two quantities, this does not pose any practical inconvenience.

\paragraph{Random Effect Selection Indicators}

For each random effect selection indicator $d_j$, $j = 1, \dots, p$, each $d_j$ once again takes on the value 0 or 1. Analogous to the main effect case, given prior $P(d_j = k) = p_{d,k}$, we have $P(d_j = k \mid \cdot) \propto \pi(\psi_j \mid d_j = k) p_{d,k}$. Once again, more concretely this comes to 

$$
P(d_j = k \mid \cdot) = \frac{\normdenszero{\xi_k}{\psi_j} p_{d,k}}{\normdenszero{\xi_k}{\psi_j} p_{d,0} + \normdenszero{\xi_k}{\psi_j} p_{d,1}}
$$

\paragraph{Random Effect Selection Probability}

Analogously to the main effect group, the final term in the main effect group is the random effect selection probability term, $P_d$. This term is entirely analogous to the $P_\gamma$ term as outlined in \ref{sec:maineffselectionprob}. For completeness, the full conditional distribution of $P_d$ is specified as follows:

$$
\pi(P_d \mid \cdot) = Beta\left(1 + \sum_{j=1}^p d_j, 1 + \left[p - \sum_{j=1}^p d_j\right]\right)
$$

\subsubsection{Error terms}

\paragraph{Global CAR Process Variance}

By definition of $\tau^2$, the full conditional is given by

$$
\pi(\tau^2 \mid \cdot) \propto \pi(\boldsymbol{\delta}_1, \dots, \boldsymbol{\delta}_N \mid \cdot) \pi(\tau^2)
$$

Because $\boldsymbol{\delta}_i(\boldmath{S}_i) \thicksim MVN(\boldmath{0}, \tau^2 [D_w(\boldmath{W}_i) - \rho \boldmath{W}_i]^{-1})$, and $\tau^2 \thicksim InvGamma(a_\tau, b_\tau)$, by basic algebraic manipulation and properties of the determinant the equation above can be expressed as 

$$
    \pi(\tau^2 \mid \cdot) \propto 
    (\tau^2)^{-S/2 - a_\tau}\exp\left(\frac{1}{\tau^2}\left[-\frac{1}{2} \sum_{i=1}^N \left(\boldsymbol{\delta}_i(\boldmath{S}_i)^T [D_w(\boldmath{W}_i) - \rho \boldmath{W}_i] \boldsymbol{\delta}_i(\boldmath{S}_i) \right)
    + b_\tau \right] \right)
$$

Thus,

$$
\pi(\tau^2 \mid \cdot) = InvGamma(a_\tau + \frac{S}{2} + 1, \frac{1}{2} \sum_{i=1}^N \left(\boldsymbol{\delta}_i(\boldmath{S}_i)^T [D_w(\boldmath{W}_i) - \rho \boldmath{W}_i] \boldsymbol{\delta}_i(\boldmath{S}_i) \right)
+ b_\tau)
$$

\paragraph{Global CAR Process Correlation}

In order to improve the computational efficiency of sampling, a grid prior was placed on the correlation of the global CAR process, $\rho$. Given a discrete set of values and corresponding probabilities, $v_1, \dots, v_K$, and $p_1, \dots, p_K$, the full conditional probability that $\rho$ is equal to $v_k$ is given by:

\begin{fleqn}
\begin{equation}
\label{eqn:rhoprop}
\begin{alignedat}{2}
    P(\rho = v_k \mid \cdot) &
\propto 
\left[\prod_{i=1}^{N} \pi(\boldsymbol{\delta}_i(\boldmath{S}_i) \mid \rho,\tau^2)\right] \pi(\rho = v_k)
\\ & \propto 
\left[\prod_{i=1}^N\left|(D_w(W_i) - v_kW_i)^{-1}\right|^{-1/2}\right]\exp\left(-\frac{1}{2\tau^2} \left[\sum_{i=1}^N \boldsymbol{\delta}_i(\boldmath{S}_i)^T (D_w(W_i) - v_kW_i) \boldsymbol{\delta}_i(\boldmath{S}_i)\right]\right)p_j
\end{alignedat}
\end{equation}
\end{fleqn}
\newcommand{\eqnrhoprop}{16 }

It follows trivially from this that

$$
P(\rho = v_k \mid \cdot) = \frac{\left[\prod_{i=1}^N\left|(D_w(W_i) - v_kW_i)^{-1}\right|^{-1/2}\right]\exp\left(-\frac{1}{2\tau^2} \left[\sum_{i=1}^N \boldsymbol{\delta}_i(\boldmath{S}_i)^T (D_w(W_i) - v_kW_i) \boldsymbol{\delta}_i(\boldmath{S}_i)\right]\right)p_j}{\sum_{t=1}^K \left[\prod_{i=1}^N\left|(D_w(W_i) - v_tW_i)^{-1}\right|^{-1/2}\right]\exp\left(-\frac{1}{2\tau^2} \left[\sum_{i=1}^N \boldsymbol{\delta}_i(\boldmath{S}_i)^T (D_w(W_i) - v_t W_i) \boldsymbol{\delta}_i(\boldmath{S}_i)\right]\right)p_t}
$$

In practice, sampling from this distribution is accomplished using the so-called Gumbel Max trick; see section \ref{sec:gumbelmax} for details and derivation. 

\paragraph{Pure Error Variance}

Denoting $\epsilon_{ij} = Y_{ij} - \mu_{ij}$, and $\boldsymbol{\epsilon} = \left[\epsilon_{11}, \dots, \epsilon_{1n_1}, \dots, \epsilon_{N1}, \dots, \epsilon_{Nn_N}\right]^T$, it follows trivially from the specification specification of the model that 

\begin{equation}
    \label{eqn:epsdist}
    \boldsymbol{\epsilon} \thicksim MVN\left(\boldmath{0}, \nu^2 \boldmath{I}\right)
\end{equation}

Combined with the fact that $\nu^2 \thicksim InvGamma(a_\nu, b_\nu)$, it trivially follows from the conjugacy of the prior that

\begin{equation}
    \label{eqn:nufc}
    \pi(\nu^2 \mid \cdot) = InvGamma(a_\nu + \frac{S}{2}, b_\nu + \sum_{i=1}^N \sum_{j = 1}^{n_i} \epsilon_{ij}^2)
\end{equation}

\subsection{Selected Genes}
Table \ref{tab:selectedgenes} shows the genes selected by the model, as well as which of the fixed- or random-effects were selected. For details on the gene composition of the combined gene sets (denoted by $*$), see Table \ref{tab:genecombos}.
\begin{table}[H]
\footnotesize
\caption{\footnotesize \textbf{Genes selected by model.} Genes selected by model, sorted by effect selection, along with which effect was selected (fixed, random, or both).}
\centering
{\rowcolors{2}{white!90!black!50}{white!70!black!50}
\begin{tabular}{|l|l|c|c|}
    \hline
    Gene & Cell Type & Fixed Effect Selected & Random Effect Selected\\
    \hline
    ITGAX & Macrophage & $\sqrt{}$ & \\
    \hline
    MAN2B1 & Macrophage & $\sqrt{}$ & \\
    \hline
    NR1H3 & Macrophage & $\sqrt{}$ & \\
    \hline
    SLC31A2 & Macrophage & $\sqrt{}$ & \\
    \hline
    GRN & Monocyte & $\sqrt{}$ & \\
    \hline
    HMOX1 & Monocyte & $\sqrt{}$ & \\
    \hline
    VCAN & Monocyte & $\sqrt{}$ & \\
    \hline
    KIR3DL2 & NK &  & $\sqrt{}$\\
    \hline
    KLRC4 & NK & $\sqrt{}$ & $\sqrt{}$\\
    \hline
    NK\_2$^*$ & NK & $\sqrt{}$ & \\
    \hline
    AQP9 & Neutrophil & $\sqrt{}$ & \\
    \hline
    BCL6 & Neutrophil & $\sqrt{}$ & \\
    \hline
    CD97 & Neutrophil &  & $\sqrt{}$\\
    \hline
    CFLAR & Neutrophil & $\sqrt{}$ & \\
    \hline
    FAM65B & Neutrophil & $\sqrt{}$ & $\sqrt{}$\\
    \hline
    Neutrophil\_1$^*$ & Neutrophil & $\sqrt{}$ & \\
    \hline
    GLT1D1 & Neutrophil &  & $\sqrt{}$\\
    \hline
    GPR97 & Neutrophil &  & $\sqrt{}$\\
    \hline
    IFITM2 & Neutrophil &  & $\sqrt{}$\\
    \hline
    KCNJ2 & Neutrophil &  & $\sqrt{}$\\
    \hline
    KIAA0247 & Neutrophil & $\sqrt{}$ & \\
    \hline
    LIMK2 & Neutrophil &  & $\sqrt{}$\\
    \hline
    MGAM & Neutrophil &  & $\sqrt{}$\\
    \hline
    NAMPT & Neutrophil &  & $\sqrt{}$\\
    \hline
    PHF21A & Neutrophil &  & $\sqrt{}$\\
    \hline
    PLXNC1 & Neutrophil & $\sqrt{}$ & $\sqrt{}$\\
    \hline
    Neutrophil\_2$^*$ & Neutrophil &  & $\sqrt{}$\\
    \hline
    STAT5B & Neutrophil &  & $\sqrt{}$\\
    \hline
    TMEM154 & Neutrophil &  & $\sqrt{}$\\
    \hline
    Plasma\_1$^*$ & Plasma & $\sqrt{}$ & $\sqrt{}$\\
    \hline
    TXNDC5 & Plasma &  & $\sqrt{}$\\
    \hline
    FYN & T &  & $\sqrt{}$\\
    \hline
    T\_1$^*$ & T & $\sqrt{}$ & \\
    \hline
    IL23A & T &  & $\sqrt{}$\\
    \hline
    PRKCH & T &  & $\sqrt{}$\\
    \hline
\end{tabular}}
\label{tab:selectedgenes}
\end{table}

\subsection{Model Trace Plots}

\begin{figure}[H]
    \centering
    \includegraphics[width=17cm]{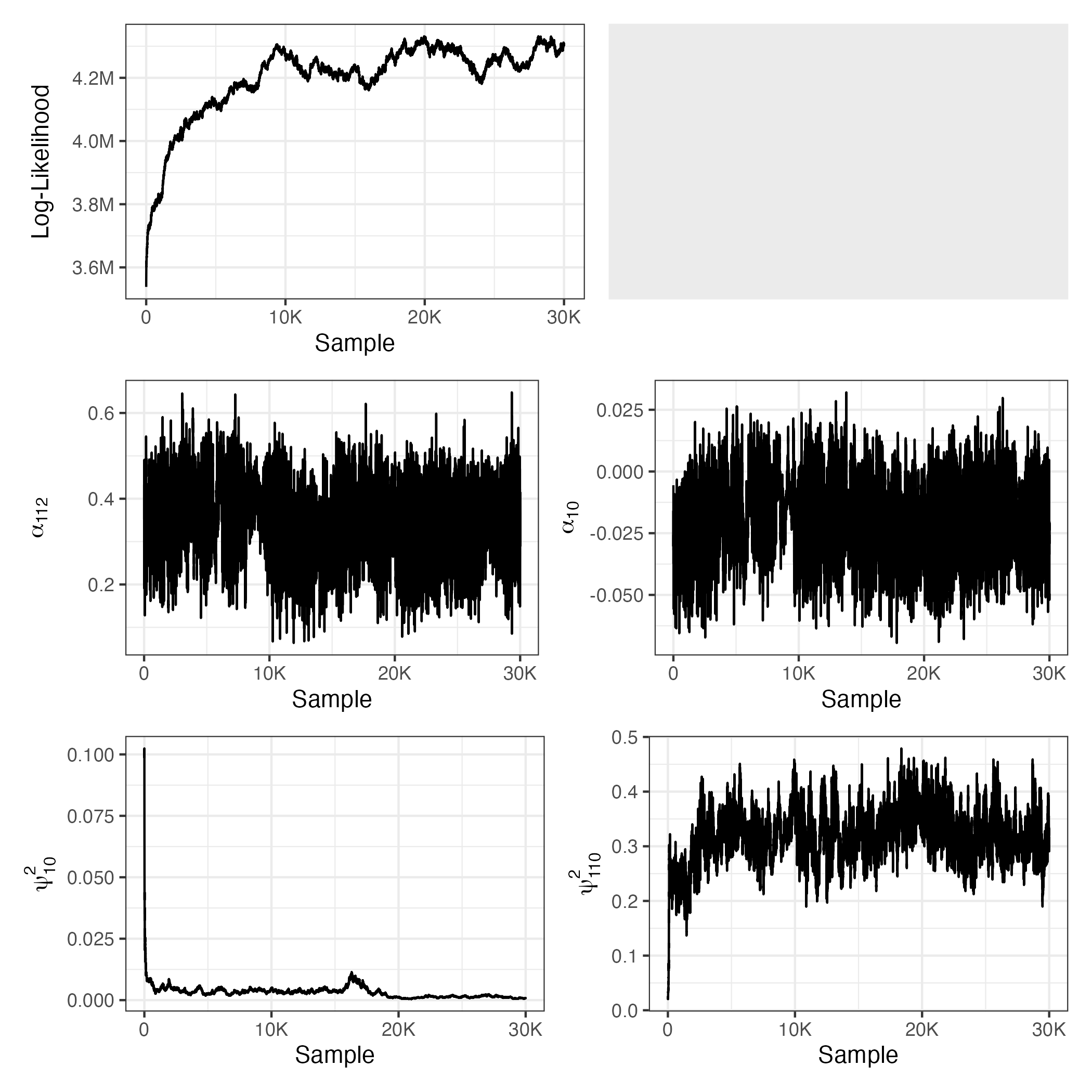}
    \caption{\textbf{Trace plots from application fitting.} Trace plots from the log-likelihood and various parameters of the application model fitting.}
    \label{fig:traceplots}
\end{figure}

The Geweke Diagnostic was used to assess convergence in the application model. Figure \ref{fig:traceplots} shows trace plots from the overall log-likelihood of the model as well as selected fixed- and random-effect parameters. 

